\def\lsim{\lower.5ex\hbox{$\; \buildrel < \over \sim \;$}}
\def\gsim{\lower.5ex\hbox{$\; \buildrel > \over \sim \;$}}
\def\fdg{\hbox{$.\!\!^\circ$}}
\documentclass{aa}
\usepackage {graphicx}
\usepackage{amssymb}
\usepackage{amsmath}
\usepackage{multirow}
\usepackage {txfonts}
\usepackage{natbib}

\begin{document}


\title{Interstellar magnetic fields in the Galactic center region}

\author{K.~Ferri\`ere\inst{1}}

\offprints{K.~Ferri\`ere}

\institute{
$^{1}$ LATT, Universit\'e de Toulouse, CNRS, 
14 avenue Edouard Belin, F-31400 Toulouse, France 
}

\date{Received  ; accepted }

\titlerunning{Interstellar magnetic fields near the Galactic center}
\authorrunning{K.~Ferri\`ere}

\abstract
{}
{We seek to obtain a picture of the interstellar magnetic field 
in the Galactic center region that is as clear and complete as possible.
}
{We review the observational knowledge that has built up over the past
25 years on interstellar magnetic fields within $\sim 200$~pc of 
the Galactic center.
We then critically discuss the various theoretical interpretations 
and scenarios proposed to explain the existing observations.
We also study the possible connections with the general Galactic
magnetic field and describe the observational situation in external galaxies.
}
{We propose a coherent picture of the magnetic field 
near the Galactic center, which reconciles some of the seemingly 
divergent views and which best accounts for the vast body of observations.
Our main conclusions are the following.
In the diffuse intercloud medium, the large-scale magnetic field 
is approximately poloidal and its value is generally close to 
equipartition with cosmic rays ($\sim 10~\mu$G), except in localized 
filaments where the field strength can reach $\sim 1$~mG.
In dense interstellar clouds, the magnetic field is approximately
horizontal and its value is typically $\sim 1$~mG.
}
{}


\maketitle

\section{\label{intro}Introduction}

The Galactic center (GC) region constitutes a very special environment,
which differs from the rest of the Galaxy both by its stellar population
and by its interstellar medium (ISM).
Here, we are not directly concerned with the stellar population,
which we tackle only to the extent that it affects the ISM,
but we are primarily interested in the ISM under its various facets.
More specifically, our purpose is to develop a comprehensive model 
of the ISM in the GC region, which can be conveniently used 
for a broad range of applications.
The particular application we personally have in mind is to study
the propagation and annihilation of interstellar positrons, which,
according to the measured annihilation emission, tend to concentrate
toward the central parts of the Galaxy 
\citep{knodlseder&jlw_05, weidenspointner&skj_06}.

In a previous paper \citep[][hereafter Paper~I]{ferriere&gj_07},
we focused on the interstellar gas in the Galactic bulge (GB),
which we defined as the region of our Galaxy interior to 
a Galactocentric radius $ \simeq 3$~kpc.
We first reviewed the current observational knowledge of its complex 
spatial distribution and physical state.
We then used the relevant observational information in conjunction with
theoretical predictions on gas dynamics near the GC
to construct a parameterized model for the space-averaged densities 
of the different (molecular, atomic and ionized) gas components.

In the present paper, we direct our attention to interstellar magnetic fields.
We start by providing a critical overview of some 25 years of
observational and interpretative work on their properties.
Since different investigation methods lead to different and sometimes
contradictory conclusions, we explore the possible sources of divergence
and try to filter out the dubious observational findings and
the questionable theoretical interpretations.
In the light of the most recent studies, we then strive to piece 
everything together into a coherent picture of interstellar magnetic fields 
in the GC region.

In Sect.~\ref{gas_distri}, we give a brief summary of the main results 
of Paper~I.
In Sect.~\ref{observ}, we review the observational evidence on
interstellar magnetic fields in the GC region, 
based on four different diagnostic tools.
In Sect.~\ref{discussion}, we provide a critical discussion of 
the various observational findings and of their current theoretical 
interpretations.
In Sect.~\ref{additional}, we examine if and how the interstellar magnetic
field in the GC region connects with the magnetic field in the Galaxy 
at large and look for additional clues from external galaxies.
In Sect.~\ref{conclu}, we present our conclusions.

\section{\label{gas_distri}Spatial distribution of interstellar gas}

\begin{figure}
\centering
\includegraphics{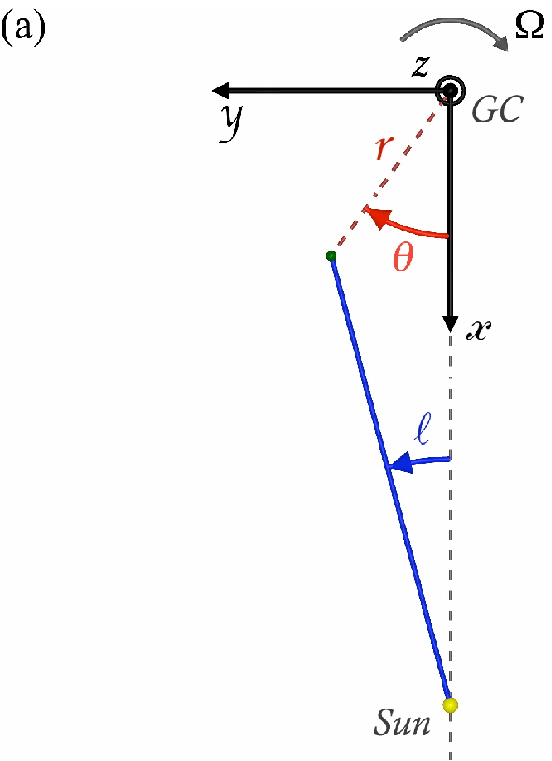}
\bigskip\bigskip\\
\includegraphics{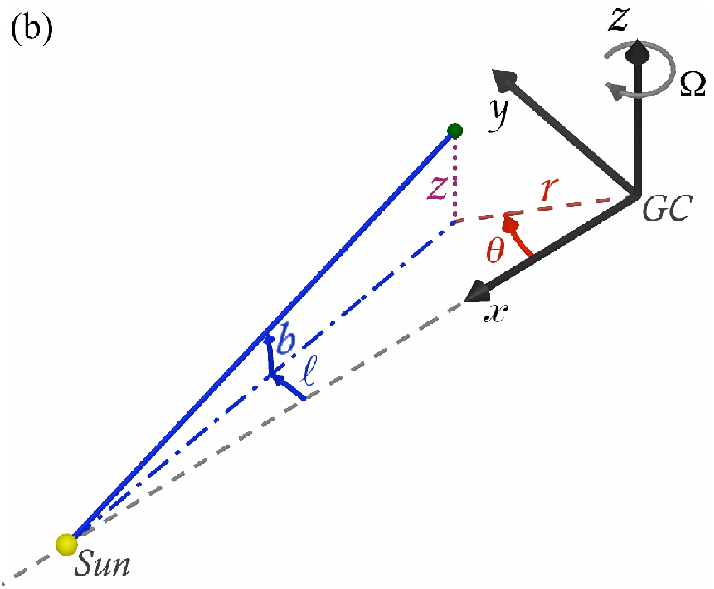}
\caption{\label{fig:coordinates}
Our $(x,y,z)$ and $(r,\theta,z)$ Galactocentric coordinate systems:
(a) face-on view from the northern Galactic hemisphere;
(b) full three-dimensional view from a point in the fourth Galactic quadrant.
}
\end{figure}

Following Paper~I, we assume that the Sun lies at a distance
$r_\odot = 8.5$~kpc from the GC,
we denote Galactic longitude and latitude by $l$ and $b$, respectively,
and employ two distinct Galactocentric coordinate systems:
(1) the cartesian coordinates $(x,y,z)$,
with $x$ measured in the Galactic plane ($b = 0^\circ$) 
along the line of sight to the Sun (positively toward the Sun), 
$y$ along the line of intersection between the Galactic plane
and the plane of the sky (with the same sign as $l$)
and $z$ along the vertical axis (with the same sign as $b$);
(2) the cylindrical coordinates $(r,\theta,z)$,
with $\theta$ increasing in the direction of Galactic rotation, 
i.e., clockwise about the $z$-axis, from $\theta = 0$ along the $x$-axis
(see Fig.~\ref{fig:coordinates}).
We further denote by $r_\perp$ the horizontal distance from the GC
projected onto the plane of the sky, which, near the GC, 
is numerically given by 
$r_\perp \simeq (150~{\rm pc}) \ (|l| / 1^\circ)$.

From the point of view of the interstellar gas, the GB possesses
two prominent structural elements: the central molecular zone (CMZ)
and the surrounding GB disk 
(also misleadingly called the H{\sc i} nuclear disk) \citep{morris&s_96}.

The CMZ is a thin sheet of gas, 
dominated at $\sim 90\%$ by its molecular component.
On the plane of the sky, it appears approximately aligned with 
the Galactic plane and displaced eastward by $\sim 50$~pc from the GC,
it extends horizontally from $y \sim -150$~pc to $y \sim +250$~pc,
and it has an {\it FWHM} thickness $\sim 30$~pc in H$_2$
and $\sim 90$~pc in H{\sc i}.
Its projection onto the Galactic plane was modeled in Paper~I
\citep[based on the observational work of][]{sawada&hhc_04}
as a $500~{\rm pc} \times 200~{\rm pc}$ ellipse, 
centered on $(x_{\rm c},y_{\rm c}) = (-50~{\rm pc},50~{\rm pc})$
and inclined clockwise by $70^\circ$ to the line of sight.
Its H$_2$ mass was estimated at $\sim 1.9 \times 10^7~M_\odot$
and its H{\sc i} mass at $\sim 1.7 \times 10^6~M_\odot$.

The GB disk is a much bigger structure, whose gas content is
$\sim 80\% - 90\%$ molecular. 
In contrast to the CMZ, it does not appear offset from the GC,
but it is noticeably tilted out of the Galactic plane
(counterclockwise by $\sim 7^\circ-13^\circ$)
and inclined to the line of sight
(near-side--down by as much as $\sim 20^\circ$).
On the plane of the sky, it extends out to a distance $\sim 1.3$~kpc 
on each side of the GC,
and it has an {\it FWHM} thickness $\sim 70$~pc in H$_2$
and $\sim 200$~pc in H{\sc i}.
According to the model described in Paper~I 
\citep[itself based on the barlike model of][]{liszt&b_80},
the GB disk has the shape of a $3.2~{\rm kpc} \times 1.0~{\rm kpc}$ ellipse,
making an angle of $51\fdg5$ clockwise to the line of sight
and featuring a $1.6~{\rm kpc} \times 0.5~{\rm kpc}$ elliptical hole 
in the middle (just large enough to enclose the CMZ).
This ``holed" GB disk has an H$_2$ mass $\sim 3.4 \times 10^7~M_\odot$ 
and an H{\sc i} mass $\sim 3.5 \times 10^6~M_\odot$.

In addition to the molecular and atomic gases, which are confined either 
to the CMZ or to the holed GB disk, the GB also contains ionized gas, 
which spreads much farther out both horizontally (beyond the boundary 
of the GB at $r \simeq 3$~kpc) and vertically (up to at least 
$|z| \simeq 1$~kpc).
This ionized gas can be found in three distinct media with completely
different temperatures.
The warm ionized medium (WIM), with $T \sim 10^4$~K, is distributed into
two components:
an extended, $\sim 2$~kpc thick disk, which continues into the Galactic disk,
and a localized, $\sim (240~{\rm pc})^2 \times 40~{\rm pc}$ ellipsoid,
nearly centered on the GC.
The hot ionized medium (HIM), with $T \sim$ a few $10^6$~K,
exists throughout the entire GB.
Finally, the very hot ionized medium (VHIM), with $T \gtrsim 10^8$~K,
is confined to a $\sim (270~{\rm pc})^2 \times 150~{\rm pc}$ ellipsoid,
centered on the GC and tilted clockwise by $\sim 20^\circ$ 
to the Galactic plane.
In Paper~I, the H$^+$ masses inside the GB were estimated
at $\sim 5.9 \times 10^7~M_\odot$ in the WIM
(with $\sim 5.8 \times 10^7~M_\odot$ in the extended disk 
and $\sim 6.0 \times 10^5~M_\odot$ in the localized ellipsoid),
$\sim 1.2 \times 10^7~M_\odot$ in the HIM
and $\sim 1.0 \times 10^5~M_\odot$ in the VHIM.

Altogether, the interstellar hydrogen content of the GB amounts to
$\sim 1.3 \times 10^8~M_\odot$, with
$\sim 5.3 \times 10^7~M_\odot$ (41\%) in molecular form,
$\sim 5.2 \times 10^6~M_\odot$ (4\%) in atomic form
and $\sim 7.1 \times 10^7~M_\odot$ (55\%) in ionized form.
Furthermore, if helium and metals represent, respectively, 
40\% and 5.3\% by mass of hydrogen, the above hydrogen masses
have to be multiplied by a factor 1.453 to obtain the total 
interstellar masses.

The broad outlines of the observed structure of the interstellar GB 
can be understood in terms of the theoretical properties of closed
particle orbits in the gravitational potential of the Galactic bar.
Basically, outside the bar's inner Lindblad resonance (ILR),
particles travel along $x_1$ orbits, which are elongated along the bar, 
whereas inside the ILR, they travel along $x_2$ orbits, 
which are elongated perpendicular to the bar.
In the presence of hydrodynamical (e.g., pressure and viscous) forces, 
the interstellar gas does not strictly follow closed particle orbits.
Instead, it gradually drifts inward through a sequence of
decreasing-energy orbits, and near the ILR it abruptly switches
from the high-energy $x_1$ orbits to the lower-energy $x_2$ orbits.
The area covered by the $x_1$ orbits forms a truncated disk,
which can naturally be identified with our holed GB disk, 
and, inside the hole, the area covered by the $x_2$ orbits
forms a smaller disk, which can be identified with the CMZ.
Although the two families of particle orbits are in theory orthogonal 
to each other,
hydrodynamical forces in the interstellar gas smooth out the $90^\circ$
jump in orbit orientation at the ILR, so that the CMZ actually leads 
the bar by less than $90^\circ$.

\section{\label{observ}Observational overview of interstellar magnetic fields}

The story begins in the 1980s, with the discovery of systems of radio
continuum filaments running nearly perpendicular to the Galactic plane.
The most striking of these systems is the bright radio arc 
(known as the Radio Arc or simply the Arc) crossing the plane near Sgr~A, 
at $l \simeq + 0\fdg18$ \citep{yusef&mc_84}.
The Radio Arc is composed of a unique set of a dozen bundled filaments,
which appear long ($\sim 40$~pc), narrow ($\sim 1$~pc)
and remarkably regular and straight.
On the other side of the GC, a fainter radio filament seems to emanate 
from Sgr~C, at $l \simeq - 0\fdg56$ \citep{liszt_85}.
The bright Radio Arc and the fainter Sgr~C filament lie vertically 
at the feet of the eastern and western ridges, respectively, 
of the $\sim 1^\circ$ $\Omega$-shaped radio lobe structure 
observed above the GC \citep{sofue&h_84}.  

The observed morphology of the radio filaments suggests that 
they trace the orientation of the local interstellar magnetic field 
\citep{yusef&mc_84,morris_90}.
This view has been borne out by radio polarization measurements, 
starting with those of \cite{inoue&ttk_84} and
\cite{tsuboi&iht_85,tsuboi&iht_86} in the pair of polarized radio lobes 
(or plumes) that make up the northern and southern extensions of 
the Radio Arc.
The authors assumed optically thin synchrotron emission and corrected
the measured polarization position angles for Faraday rotation, 
whereupon they found that the transverse (to the line of sight) 
magnetic field ${\bf B}_\perp$ is indeed oriented along the axis 
of the plumes.

Thus, the orientation of the radio filaments provides evidence that 
the surrounding interstellar magnetic field is approximately normal to 
the Galactic plane.
In addition, their slight outward curvature indicates that the magnetic 
field does not remain vertical at large distances from the plane, 
but that it turns instead to a more general poloidal geometry
\citep{morris_90}.

\cite{tsuboi&iht_86} derived a crude estimate for the magnetic field strength
$B$ in the polarized radio lobes, based on their measured (supposedly
synchrotron) radio intensity and on the assumption of energy
equipartition between the magnetic field and the energetic particles.
They thereby obtained a magnetic field strength of several $10~\mu$G,
compatible with the line-of-sight field $\sim 10~\mu$G deduced from 
Faraday rotation measures \citep{tsuboi&iht_85}. 
\cite{yusef&m_87b}, on the other hand, estimated the equipartition 
field strength in the radio filaments at $\sim 0.2$~mG.

\cite{yusef&m_87b} also provided an independent magnetic field strength 
estimate for the filaments of the Radio Arc with the help of a simple 
dynamical argument \citep[see also][]{yusef&m_87c}.
The fact that the filaments remain nearly straight all along their length
and that they pass through the Galactic plane with little or no bending
suggests that their magnetic pressure, $P_{\rm mag}$, is strong enough 
to withstand the ram pressure of the ambient interstellar clouds, 
$P_{\rm ram}$. The condition $P_{\rm mag} \gtrsim P_{\rm ram}$, together with
a conservative estimation of $P_{\rm ram}$, then leads to the stringent 
requirement that $B \gtrsim 1$~mG inside the radio filaments.

A similar dynamical argument can be applied to the less prominent
nonthermal radio filaments (NRFs) discovered later on. 
Going one step further, it is then possible to estimate the magnetic
field strength outside the NRFs by invoking pressure balance.
As argued by \cite{morris_90}, the external ISM must supply a confining
pressure for the NRFs.
This pressure cannot be of thermal origin, because even the very hot gas,
which has the highest thermal pressure, falls short by a factor $\sim 30$.
The confining pressure must, therefore, be of magnetic origin,
which means that the mG field inferred to exist inside the NRFs 
must also prevail outside.
According to this argument, the reason why the NRFs stand out in the radio maps
is not because they have a stronger magnetic field than their surroundings,
but because they contain more relativistic electrons.
To quote \cite{morris_90}, the NRFs ``are illuminated flux tubes within 
a relatively uniform field''.

To sum up, the picture emerging at the end of the 1980s 
for the interstellar magnetic field within $\sim 70$~pc of the GC 
is that of a pervasive, poloidal, mG field.
The magnetic energy contained in this mG field is as high as
$\sim (1-2) \times 10^{54}$~ergs inside 70~pc 
\citep{morris_90},\footnote{
At the present time, NRFs are observed over a larger area,
extending $\sim 300$~pc along the Galactic plane and $\sim 150$~pc 
in the vertical direction \citep{larosa&nlk_04}. 
The magnetic energy of a space-filling mG field in this larger region 
is a huge $\sim 10^{55}$~ergs.
}
which is equivalent to the energy released by $\sim (1000 - 2000)$ 
supernova explosions, and which is roughly comparable to the kinetic
energy associated with Galactic rotation in the considered region, 
while being significantly larger than both the kinetic energy 
in turbulent motions and the thermal energy of the very hot gas.

The simple view of a pervasive, poloidal, mG field 
gained wide acceptance, 
until a variety of subsequent observations gradually called it into question.
In the remainder of this section, we review the relevant observations
that have either given a new twist to the simple picture described above
or otherwise contributed to enhance our understanding of 
the interstellar magnetic field near the GC.

\subsection{\label{radio}Radio continuum observations}

After the initial discovery of \cite{yusef&mc_84}, numerous NRFs were
identified in the GC region.
Although the Radio Arc is clearly a unique structure, the other NRFs 
share a number of observational characteristics with the Radio Arc's 
filaments.
\cite{morris_96} drew up the inventory of all the GC NRFs known at the time
and summarized their distinctive properties, i.e., the properties 
that make them unique to the GC region.
In brief, the GC NRFs are a few to a few tens of parsecs long and 
a fraction of a parsec wide; they look straight or mildly curved
along their entire length; they run roughly perpendicular to 
the Galactic plane; and their radio continuum emission is both 
linearly polarized and characterized by a spectral index consistent 
with synchrotron radiation.

To this list, we should add that the GC NRFs have equipartition or 
minimum-energy field strengths\footnote{
The minimum-energy state corresponds almost, but not exactly, to energy 
equipartition between the magnetic field and the energetic particles 
\citep{miley_80}, which, in turn, differs from pressure equipartition
by a factor $\simeq 3$.
Both equipartition and minimum-energy field strengths are difficult 
to estimate, because they depend on a number of parameters whose values
are quite uncertain.
In particular, they depend on the proton-to-electron energy ratio
(often set to unity in this context), on the lower and upper cutoff
frequencies of the synchrotron spectrum, on the spectral index,
and on the line-of-sight depth of the filaments (usually taken equal 
to their plane-of-sky width).
}
of several $10~\mu$G, typically $\sim (50 - 200)~\mu$G
\citep[and references therein]{anantharamaiah&peg_91,larosa&nlk_04}.
Since they appear to be magnetically dominated, it is likely
that these values underestimate the true field strengths.
This could perhaps partly explain why the equipartition/minimum-energy 
field strengths are systematically lower than the mG field strength 
deduced from the dynamical condition $P_{\rm mag} \gtrsim P_{\rm ram}$.
Even so, while there is general agreement that the GC NRFs 
represent magnetic flux tubes lit by synchrotron-emitting electrons,
the question of how strong their intrinsic magnetic field really is 
remains open to debate.
Another unsettled question is whether their strong field is limited 
to their interiors or representative of the GC region as a whole.

Amongst all the NRFs cataloged by \cite{morris_96}, the one dubbed the Snake, 
located at $l \simeq - 0\fdg90$ ($r_\perp \simeq 135$~pc), is atypical 
by its morphology which exhibits two marked kinks along its length 
\citep{gray&ceg_91}.
This kinked shape probably reveals that the magnetic field 
inside the Snake is weaker than the mG field required to withstand 
the ambient ram pressure. 
Incidentally, a weaker field is easier to reconcile with 
the minimum-energy field strength $\simeq 90~\mu$G derived by
\cite{gray&nec_95}.
Since the Snake lies at a greater projected radius than the other NRFs, 
a simple way to explain its weaker field is to invoke an overall decrease
in the interstellar magnetic field strength outside a radius
$\gtrsim 100$~pc.

Another interesting NRF, discovered somewhat later by \cite{lang&akl_99}, 
is the so-called Pelican, located at $l \simeq - 1\fdg15$ 
($r_\perp \simeq 172$~pc).
It, too, is noticeably kinked, but what distinguishes it most from 
the other NRFs is its orientation parallel to the Galactic plane.
Its equipartition field strength is $\simeq 70~\mu$G, 
and the measured polarization position angles, corrected for Faraday rotation, 
confirm that its intrinsic ${\bf B}_\perp$ is everywhere aligned with
its long axis.
The Pelican's anomalous features provide a hint that the interstellar
magnetic field outside a certain radius, say, $\sim 200$~pc,
is not only less intense than inside $\sim 100$~pc, but also oriented 
approximately parallel to the Galactic plane.
In fact, 
the actual situation is probably more complex,
as an NRF oriented at $45^\circ$ to the Galactic plane was found
much closer to the GC, at $l \simeq - 0\fdg68$ ($r_\perp \simeq 102$~pc)
\citep{larosa&nlk_04}.

\cite{yusef&hc_04} presented a sensitive 20-cm continuum survey 
of the GC region $(-2^\circ < l < 5^\circ$, $-40' < b < 40')$,
which resulted in a new catalog of more than 80 linear radio filaments, 
including all the previously well-established NRFs as well as many new 
good candidates.
The authors also drew a schematic diagram of the distribution of all 
the radio filaments, which puts them into perspective,
with their respective positions, orientations, sizes and shapes 
(see their Fig.~29).
A cursory look at the diagram confirms that the vast majority 
of filaments are nearly straight and that they have a general 
tendency to run along roughly vertical axes.
However, upon closer inspection, it appears that only the longer
filaments strictly follow this tendency.
The shorter filaments exhibit a broad range of orientations, 
with only a loose trend toward the vertical.
On the other hand, \cite{yusef&hc_04}'s diagram does not point to any 
obvious correlation between either the orientation or the shape 
(overall curvature, presence of kinks) of the filaments and 
their projected distance from the GC.

The high-resolution, high-sensitivity 330~MHz imaging survey of 
\cite{nord&lkh_04} leads to similar conclusions.
While the brightest NRFs (with the exception of the Pelican)
tend to align along the vertical, the newly discovered fainter NRFs 
(or candidate NRFs) have more random orientations, with a mean angle
to the vertical $\simeq 35^\circ \pm 40^\circ$.
Here, too, some of the strongly inclined or nearly horizontal NRFs 
lie much closer to the GC than the Pelican.

The GC region was also imaged at 
74~MHz and 330~MHz by \cite{larosa&bsl_05}.
Their high-resolution 74~MHz image reveals, aside from discrete emission
and thermal absorption features, a $6^\circ \times 2^\circ$ source of
diffuse nonthermal emission centered on the GC 
(note that the NRFs detected at higher frequencies are resolved out 
in this image).
The source of diffuse emission is also clearly visible 
in the lower-resolution 330~MHz image.
\cite{larosa&bsl_05} estimated the integrated 74~MHz and 330~MHz 
flux densities and the spectral index of the (supposedly synchrotron) 
diffuse emission, and from this they derived a minimum-energy field strength 
(on spatial scales $\gtrsim 5$~pc) 
$\simeq (6~\mu{\rm G})~(\phi / f)^{2/7}$, 
where $\phi$ is the proton-to-electron energy ratio 
and $f$ the filling factor of the synchrotron-emitting gas.
They noted that even the combination of extreme values $\phi \simeq 100$ 
and $f \simeq 0.01$ leads to a minimum-energy field strength
$\lesssim 100~\mu$G.
Although the large-scale magnetic field could easily be stronger 
than the minimum-energy value, \cite{larosa&bsl_05} advanced a number of 
cogent arguments against a large-scale field as strong as 1~mG.
With a 1~mG field, they argued, the radiative lifetime of 
synchrotron-emitting electrons would only be $\sim 10^5$~yr,
``shorter than any plausible replenishment timescale''.
In addition, the measured synchrotron flux would imply a cosmic-ray 
electron energy density of only $\sim 0.04$~eV~cm$^{-3}$, which is
about 5 times lower than in the local ISM, whereas several lines 
of evidence (in particular, from the observed Galactic diffuse
$\gamma$-ray emission) point to comparable values.

Because of the uncertainties inherent in the equipartition/minimum-energy 
assumption made to exploit radio synchrotron emission data,
there have been attempts to derive the relativistic-electron density 
by separate means, for instance, based on X-ray emission spectra.
\cite{yusef&wml_05} detected X-ray emission from the radio filament 
G359.90$-$0.06, which they tentatively attributed to inverse Compton 
scattering of far-infrared photons from dust by the relativistic
electrons that produce the radio synchrotron emission.
Relying on dust thermal emission observations to constrain the dust
parameters, they were able to infer the relativistic-electron density 
from the X-ray flux and then the magnetic field strength from the radio 
synchrotron flux.
In this manner, they estimated the field strength in the considered NRF 
at $\sim (30-130)~\mu$G. Quite unexpectedly, this loose range of values
falls below the equipartition field strength, estimated here at 
$\sim (140-200)~\mu$G.

\subsection{\label{Faraday}Faraday rotation measurements}

The first radio polarization studies of the Radio Arc by
\cite{inoue&ttk_84} and \cite{tsuboi&iht_85} included measurements
at four different frequencies around 10~GHz ($\lambda = 3$~cm),
which enabled the authors to examine the wavelength-dependence 
of both the polarization position angle and the polarization degree 
of the incomig radiation.
\cite{inoue&ttk_84} focused on two strongly polarized regions 
along the Radio Arc and its extensions, 
which roughly coincide with the peaks in polarized intensity 
in the southern and northern radio lobes.
They derived unusually large rotation measures (RMs)
$\simeq -1660~{\rm rad~m^{-2}}$ for the southern region (Source~A)
and $\simeq +800~{\rm rad~m^{-2}}$ for the northern region (Source~B).
With their assumed values of the thermal-electron density, 
$n_{\rm e} \sim 100~{\rm cm}^{-3}$, and of the path length through
the Faraday-rotating medium, $L \sim 5$~pc, the derived RM of 
Source~A translates into a line-of-sight magnetic field 
$B_\parallel \sim -4~\mu$G.\footnote{
The exact formula for the rotation 
measure is ${\rm RM} = 0.81 \ \int n_{\rm e} \ B_\parallel \ ds$,
with ${\rm RM}$ expressed in ${\rm rad~m^{-2}}$, 
$n_{\rm e}$ in ${\rm cm^{-3}}$, $B_\parallel$ in $\mu{\rm G}$
and $s$ in ${\rm pc}$.
}\footnote{
By convention in the Faraday rotation community,
a positive (negative) value of $B_\parallel$ corresponds to 
a magnetic field pointing toward (away from) the observer.
}
However, if Faraday rotation occurs within the synchrotron-emitting 
region itself (rather than in the foreground ISM), 
as suggested by the steep decrease in the polarization degree 
with increasing wavelength,
then the implied value of $B_\parallel$ is in fact twice higher.

\cite{tsuboi&iht_85}, who mapped a whole $0\fdg6 \times 1\fdg2$ area
straddling the Radio Arc, obtained positive RMs almost everywhere 
in the northern lobe, with a maximum value $\simeq +600~{\rm rad~m^{-2}}$ 
near the peak in polarized intensity, and negative RMs around the peak 
in polarized intensity of the southern lobe, with a minimum value 
$\simeq -2000~{\rm rad~m^{-2}}$ and an average value 
$\simeq -1500~{\rm rad~m^{-2}}$.
They also found negative RMs near the northeastern edge 
of the northern lobe (a marginal detection, later confirmed by 
\cite{tsuboi&iht_86}) and positive RMs in the extended, patchy tail 
of the southern lobe (not confirmed by \cite{tsuboi&iht_86}).
For the region around the peak of the southern lobe, they assumed
$n_{\rm e} \sim 30~{\rm cm}^{-3}$ and $L \sim 5$~pc,
thereby arriving at $B_\parallel \sim -10~\mu$G.
Again, this value should be multiplied by a factor of 2 
if Faraday rotation is internal to the source of synchrotron emission.

Evidently, the line-of-sight fields inferred here from Faraday RMs 
can only be regarded as rough estimates, given the large 
uncertainties in the nature of the Faraday-rotating medium, 
in its location with respect to the GC (can it truly be localized 
to the GC?) and with respect to the synchrotron-emitting region
(are both domains spatially coincident or separate?),
in its line-of-sight depth, and in its thermal-electron density.
But even so, the above studies bring to light a clear magnetic field 
reversal between the northern and southern radio lobes
as well as two possible reversals near the northeastern edge 
of the northern lobe and between the peak and the tail
of the southern lobe.

To explain the main field reversal, \cite{tsuboi&iht_86} invoked the presence
of a magnetic flux tube running through the Galactic plane and bent
somewhere near the midplane in such a way that the magnetic field 
globally points toward (away from) the observer above (below) the midplane.
This bending could be due either to a local interaction with a nearby
molecular cloud or to Galactic rotation.
The secondary reversals, if confirmed, could be explained by 
a propagating Alfv\'en wave, which would also account for the waving 
pattern observed in the distribution of ${\bf B}_\perp$.
Let us emphasize that \cite{tsuboi&iht_86}'s interpretation does not
contradict the notion that the NRFs composing the Radio Arc are nearly
straight, as the bending implied by $|B_\parallel| \sim 10~\mu$G
is indeed very modest if $B \sim 1~$mG.
One might even see here an independent piece of evidence in support of a mG
field, or at least, of a field $\gg 10~\mu$G -- unless the bending
fortuitously occurs in the vertical plane containing the line of sight.

Further, high-resolution observations at $\lambda \, 6$~cm and 
$\lambda \, 20$~cm
by \cite{yusef&m_87b} toward a segment of the Radio Arc comprising
Source~A of \cite{inoue&ttk_84} yielded negative RMs down to 
$\simeq -5500~{\rm rad~m^{-2}}$ in the vicinity of Source~A
and positive RMs up to $\simeq +350~{\rm rad~m^{-2}}$ farther south.
Similar observations by \cite{yusef&m_88} toward a small area
of the northern lobe comprising Source~B yielded positive RMs up to
$\simeq +1450~{\rm rad~m^{-2}}$.
Both sets of RMs are in reasonably good agreement with the results 
obtained at $\lambda \, 3$~cm by \cite{inoue&ttk_84} and \cite{tsuboi&iht_85}.
However, in contrast to these authors, \cite{yusef&m_88} attributed
their observed RMs to a combination of internal and foreground Faraday
rotation, while \cite{yusef&m_87b}, noticing locations with large RMs and 
only weak depolarization, suggested that Faraday rotation takes place 
outside the filaments, possibly in a helical magnetic structure 
surrounding them and having 
$L \sim 0.3$~pc, $n_{\rm e} \sim 2200~{\rm cm}^{-3}$ and 
$B_\parallel \sim -10~\mu$G. 

On the other hand, \cite{tsuboi&kkh_95} observed the Radio-Arc +
polarized-lobes complex at 42.5~GHz ($\lambda = 0.7$~cm),
a higher frequency at which Faraday rotation and Faraday depolarization 
should be nearly negligible.
Not only did they confirm the RMs derived by \cite{inoue&ttk_84} and 
\cite{tsuboi&iht_85}, but they also made their case for internal Faraday
rotation stronger.
Clearly, the contradictory interpretations proposed by different authors 
for the Radio Arc and polarized lobes attest to the difficulty of locating 
the Faraday screen with respect to the synchrotron source.

Let us now turn to the Snake (see Sect.~\ref{radio}), which was 
observed by \cite{gray&nec_95} at five different frequencies ranging 
between 0.84~GHz and 8.64~GHz.
By analyzing the wavelength-dependence of its polarization properties,
they found that the Snake experiences both internal Faraday rotation,
with Faraday depth $\sim +1400~{\rm rad~m^{-2}}$, and foreground 
Faraday rotation, with RM $\sim +5500~{\rm rad~m^{-2}}$.
For the external Faraday-rotating medium, they estimated
$n_{\rm e} \sim 10~{\rm cm}^{-3}$, $L \sim 100$~pc and thus
$B_\parallel \sim 7~\mu$G, while for the Snake itself, they mentioned
that the measured internal Faraday depth could be reproduced with
$n_{\rm e} \sim 10~{\rm cm}^{-3}$, $L \sim 1$~pc (approximate width
of the Snake) and $B_\parallel \sim 90~\mu$G (minimum-energy field 
strength; see Sect.~\ref{radio}).
Note that the latter value is not particularly relevant, 
as there is no reason why $B_\parallel$ (which is probably only a minor 
component of ${\bf B}$) would equal the minimum-energy field strength.

Another instructive NRF is G359.54+0.18, observed by \cite{yusef&wp_97}
at $\lambda \, 3.6$~cm and $\lambda \, 6$~cm.
The authors obtained RMs down to $\simeq -4200~{\rm rad~m^{-2}}$,
with typical values in the range $\sim -3000~{\rm rad~m^{-2}}$ 
to $-1500~{\rm rad~m^{-2}}$. They argued that Faraday rotation occurs
in a foreground screen, which they suggested could be the $10^8$~K gas 
present in the inner 100~pc of the Galaxy. 
With $n_{\rm e} \sim 0.03~{\rm cm}^{-3}$ \citep{yamauchi&kkk_90},
this gas would require $B_\parallel \sim -1$~mG to explain the typical
RM values.
And if, as suggested by the small-scale RM variations along 
the filament, the $10^8$~K gas is clumped with a filling factor 
$\Phi \sim 0.01$, the values of $n_{\rm e}$ and $B_\parallel$ 
should both be divided by $\sqrt{\Phi} \sim 0.1$,
which would make $|B_\parallel|$ even more unrealistically high. 
However, we note that the $10^8$~K gas encloses but a small fraction 
of the interstellar free electrons on the line of sight, 
the vast majority of them residing in the warm $\sim 10^4$~K phase 
of the ISM (see Paper~I).
Accounting for all the interstellar free electrons, supposed to be distributed
according to \cite{cordes&l_02}'s model, we find that their integrated
density from the GC to Galactic radius $r$ rises steeply with $r$ out to
$\sim 200$~pc, with $\int_{0}^{200\,{\rm pc}} n_{\rm e} \ ds \simeq 
1230~{\rm cm^{-3}~pc}$, then gradually levels off to $\int_{0}^{r_\odot}
n_{\rm e} \ ds \simeq 1960~{\rm cm^{-3}~pc}$.
It then follows that the above typical RMs translate into 
$\overline{B}_\parallel \sim -\,(1-2)~\mu$G, 
where the overbar denotes a line-of-sight average weighted by $n_{\rm e}$,
i.e., in the present context, a line-of-sight average heavily weighted 
toward the region $r \lesssim 200$~pc.
Obviously, $\overline{B}_\parallel$ could significantly underestimate
the typical local $B_\parallel$ in this region 
if the magnetic field reverses along the line of sight.

For the Pelican, observed by \cite{lang&akl_99} at $\lambda \, 3.6$~cm,
$\lambda \, 6$~cm and $\lambda \, 20$~cm,
the measured RMs vary smoothly from $\simeq -500~{\rm rad~m^{-2}}$ 
at the western end to $\simeq +500~{\rm rad~m^{-2}}$ at the eastern end,
with a peak value $\simeq -1000~{\rm rad~m^{-2}}$.
Faraday rotation is most probably external, as implied by the particularly
high polarization degrees. 
If so, the fact that the Pelican has somewhat lower RMs than the other NRFs
could indicate that it lies slightly closer to the Sun, 
and the sign reversal in RM could reflect the presence of a magnetic 
perturbation in front of the Pelican, with field lines bent out
of the plane of the sky.
Otherwise, the authors made no attempt to convert their measured RMs 
into estimates for $B_\parallel$.

More RM studies toward other NRFs have been carried out in the last decade
\cite[e.g.,][]{lang&me_99}.
The derived RMs typically range from a few hundred to a few thousand
${\rm rad~m^{-2}}$. 
When these RMs are used to estimate $B_\parallel$ near the GC, 
values of a few $\mu$G are usually obtained.
As already mentioned above, these values are extremely uncertain, 
due to our poor knowledge of the precise free-electron density
distribution along the considered line of sight.
Furthermore, their exact significance remains unclear,
first because they refer to line-of-sight averages (with possible 
cancellations if $B_\parallel$ reverses sign) through ionized regions only, 
and second because $B_\parallel$ is likely to represent but a small 
component of the total magnetic field.
In this respect, the inferred values of $B_\parallel$ might tell us 
more about the rigidity of the interstellar magnetic field near the GC 
than about its strength.

RM studies also provide valuable information on the magnetic field
geometry near the GC. 
By collecting all the available RMs toward NRFs within $1^\circ$ 
of the GC, \cite{novak&crg_03} were able to uncover
a definite pattern in the sign of RM, such that 
${\rm RM} > 0$ in the quadrants $(l>0,b>0)$ and $(l<0,b<0)$ and 
${\rm RM} < 0$ in the quadrants $(l>0,b<0)$ and $(l<0,b>0)$.
This pattern, they explained, could be understood as the result of 
an initially axial magnetic field (pointing north) being sheared out 
in the azimuthal direction by the Galactic differential rotation 
(dense gas near the Galactic plane tends to rotate faster than 
diffuse gas higher up).
The RM results described in this section show that the actual situation 
is not as clear-cut, as we came across several filaments that exhibit
both signs of RM in the same quadrant.
For the filaments of the Radio Arc, one sign clearly dominates,
and the other sign could be attributed to an Alfv\'en wave
traveling along the filaments \citep{tsuboi&iht_86}.
For the Pelican (which lies slightly outside $1^\circ$ of the GC), 
both signs are equally important, and the sign reversal could be 
attributed to a foreground magnetic perturbation (see above).

\cite{novak&crg_03}'s conclusions are not supported by the recent work
of \cite{roy&rs_05} and \cite{roy&rs_08}, who considered the more 
extended area $(|l|<6^\circ,|b|<2^\circ)$. 
\cite{roy&rs_05} observed 59 background extragalactic sources through 
this area, at $\lambda \, 3.6$~cm and $\lambda \, 6$~cm, 
and obtained RMs in the range $\simeq -1180~{\rm rad~m^{-2}}$ 
to $+4770~{\rm rad~m^{-2}}$.
\cite{roy&rs_08} remarked that these RMs are predominantly positive, 
with a mean value $\simeq +413~{\rm rad~m^{-2}}$,
and they found no evidence for a RM sign reversal either 
across the rotation axis or across the midplane.
As discussed by the authors, this observed RM distribution
is consistent with either the large-scale Galactic magnetic field 
having a bisymmetric spiral configuration or the magnetic field in 
the central region of the Galaxy being oriented along the Galactic bar.
Interestingly, the 4 extragalactic sources of the sample that fall closer 
than $1^\circ$ of the GC conform neither to \cite{novak&crg_03}'s pattern 
(only 2 have the expected RM sign) nor to the general pattern of the more 
extended area (only 2 have ${\rm RM} > 0$).

It is clear that a much broader sample of RM data toward the GC
would be needed to draw firm conclusions on the magnetic field geometry 
near the GC.
It would also be necessary to get a better handle on how much 
of the observed RMs can truly be attributed to the GC region.

\cite{roy&rs_08} argued that their RMs have negligible
contributions from the sources themselves,
they estimated the contribution from the Galactic disk to be small
and, based on the measured RMs of 7 pulsars located within the survey area 
(with distances between 1.5~kpc and 7.7~kpc and with a mean RM of only 
$\simeq (-7 \pm 46)~{\rm rad~m^{-2}}$), they ruled out the possibility 
that a single high-RM object (H{\sc ii} region or supernova remnant) 
could bias their entire sample.
From this, they concluded that their RMs arise mainly from the central 
$\sim 2$~kpc of the Galaxy.
Relying on \cite{bower&bs_08}'s observations, they further estimated 
the free-electron column density across the central 2~kpc at 
$\sim 800~{\rm cm^{-3}~pc}$, which they combined with the mean RM 
$\simeq +413~{\rm rad~m^{-2}}$ to derive a mean
$\overline{B}_\parallel \sim 0.6~\mu$G.
Again, this value by itself is not very meaningful, as it represents 
an $n_{\rm e}$-weighted average, 
over a $12^\circ \times 4^\circ$ cone opening up from the Sun toward the GC,
of a quantity that undergoes repeated sign reversals.
More useful is the estimate of the turbulent (or random) component 
of $B_\parallel$. 
From the RM structure function and the r.m.s RM of their sample, 
\cite{roy&rs_08} estimated $\delta B_\parallel \sim 6~\mu$G 
in the region $r \lesssim 1$~kpc and $\delta B_\parallel \sim 20~\mu$G 
in the region $r \lesssim 150$~pc.

\subsection{\label{IR_Smm}Infrared and sub-millimeter polarimetry}

Interstellar dust grains tend to spin about their short axes and
to align the latter along the local magnetic field. 
As a result, the dust thermal emission, at far-infrared (FIR) and 
sub-millimeter (Smm) wavelengths, is linearly polarized perpendicular 
to the ambient magnetic field.
It then follows that FIR/Smm polarimetry makes it possible to map out 
the direction of the interstellar magnetic field on the plane of the sky,
in regions of strong dust emissivity, i.e., in high-density regions
\citep[e.g.,][]{hildebrand_88}.
The first successful applications of this technique to the GC area
were by \cite{aitken&brb_86,aitken&brb_91}.

\cite{davidson_96} reviewed the existing FIR polarization measurements
toward dense regions located within $\sim 100$~pc of the GC.
In the four regions that she discussed (certainly in the Circumnuclear
Disk\footnote{
The Circumnuclear Disk is a compact torus of neutral 
(mainly molecular) gas and dust surrounding Sgr~A West 
(the H{\sc ii} region centered on the point-like source Sgr~A$^*$) 
and extending between $\sim 1.5$~pc and 7~pc from the GC
\citep{morris&s_96, mezger&dz_96}.
} 
(CND) and associated Northern Streamer, in the Arched Filaments and 
in the Sickle, and possibly in Sgr~B2), the measured field direction
is roughly parallel to the Galactic plane.
\cite{davidson_96} argued that this field direction could be explained 
by the dense gas moving relative to the surrounding diffuse gas and 
either distorting the local poloidal magnetic field or dragging 
its own distorted field from another Galactic location.
She also quoted a mean magnetic field strength $\sim 6$~mG in the Arched
Filaments, as inferred from the Chandrasekhar-Fermi relation 
\citep[see][]{morris&dwd_92}.

Compared to FIR polarimetry, which probes the warmer parts of 
molecular clouds, Smm polarimetry applies to their colder parts.
Polarized Smm emission from the GC region was first detected by
\cite{novak&ddh_00}, who carried out $350~\mu$m polarimetric observations 
of three separate $2' \times 2'$ 
areas, centered on the CND and on the peaks of the M$-$0.02$-$0.07
and M$-$0.13$-$0.08 molecular clouds, respectively.
In all three areas, the measured field directions are 
more-or-less inclined to the Galactic plane.
In the CND, they are similar to the field directions given by 
FIR polarimetry;
in the curved ridge of M$-$0.02$-$0.07, they are everywhere aligned 
with the ridge, consistent with the field being compressed together 
with the gas by the expansion of Sgr~A East;
and in M$-$0.13$-$0.08, they are on average parallel to the cloud 
long axis, consistent with the field being stretched out by the tidal 
forces that gave the cloud its elongated shape.
These first results already show that the magnetic field configuration 
near the GC is governed by a combination of different factors.

A much larger area about the GC, extending 170~pc in longitude and 
30~pc in latitude, was observed in $450~\mu$m polarized emission 
by \cite{novak&crg_03}.
Their polarization map clearly shows that the magnetic field threading
molecular clouds is, on the whole, approximately parallel 
to the Galactic plane.
To reconcile the horizontal field measured in molecular clouds with 
the poloidal field traced by NRFs, \cite{novak&crg_03} suggested 
that the large-scale magnetic field near the GC is predominantly poloidal 
in the diffuse ISM and predominantly toroidal in dense regions along 
the Galactic plane, where it was sheared out in the azimuthal direction 
by the differential rotation of the dense gas.

\cite{chuss&ddd_03} reported further $350~\mu$m polarimetric observations
of the central 50~pc of the Galaxy, which led them to refine
the conclusions of \cite{novak&crg_03}.
They found that the measured field direction depends on the molecular
gas density in such a way that it is generally parallel to the Galactic 
plane in high-density regions and generally perpendicular to it 
in low-density regions.
They proposed two possible scenarios to explain their results.
In their preferred scenario, the large-scale magnetic field was 
initially poloidal everywhere, but in dense molecular clouds, where 
the gravitational energy density exceeds the magnetic energy density, 
it became sheared out into a toroidal field by the clouds' motions.
In the alternative, less likely scenario, the large-scale magnetic field 
was initially toroidal everywhere, but outside dense molecular clouds,
it became distorted into a poloidal field by winds due to supernova
explosions.
If the first scenario is correct, a characteristic field strength 
inside GC molecular clouds can be estimated by assuming that clouds 
where the field is half-way between toroidal and poloidal
(i.e., inclined by $45^\circ$ to the vertical) are those where 
gravitational and magnetic energy densities are equal. 
A crude estimation of the gravitational energy then yields
a characteristic field strength $\sim 3$~mG inside molecular clouds.\footnote{
This rough estimate of $B$ was necessarily going to be larger 
than the mG estimate derived for the diffuse ISM 
by \cite{yusef&m_87c,yusef&m_87b} 
from the dynamical condition $P_{\rm mag} \gtrsim P_{\rm ram}$ 
(see beginning of Sect.~\ref{observ}), since the orbital velocity 
of molecular clouds in the Galactic gravitational potential 
($\sim 150~{\rm km~s}^{-1}$) exceeds their turbulent velocity 
($\sim 15~{\rm km~s}^{-1}$).
}

While dust grains {\it emit} polarized thermal radiation at FIR/Smm 
wavelengths, they {\it absorb} starlight at optical and near-infrared 
(NIR) wavelengths.
Optical starlight from the GC region becomes completely extinct before
reaching us, but NIR starlight suffers only partial extinction, such that
it reaches us linearly polarized in the direction of the magnetic field 
(the dust-weighted average field 
along the line of sight to the observed stars).
For this reason, NIR polarimetry toward the GC has become a new tool 
to trace the average magnetic field direction on the plane of the sky,
in the GC region.

\cite{nishiyama&thk_08} obtained a NIR polarization map of 
a $50~{\rm pc} \times 50~{\rm pc}$ area centered on the GC.
Compared to earlier NIR polarimetric observations toward the GC
\citep[e.g.,][]{eckart&ghs_95, ott&eg_99}, they were able, 
for the first time, to separate out the contribution from foreground dust
and to isolate the polarization arising within $\sim (1-2)$~kpc 
of the GC.
Their inferred distribution of polarization position angles exhibits 
a strong peak in a direction nearly parallel to the Galactic plane,
in good agreement with the results of FIR/Smm polarimetry.
However, in contrast to \cite{chuss&ddd_03}, \cite{nishiyama&thk_08} found 
no indication that the magnetic field direction depends on gas density -- 
the field appears to be everywhere horizontal, including in the diffuse ISM.

These first NIR results are preliminary, but they demonstrate 
the potential of NIR polarimetry to probe the GC magnetic field.
This potential will probably prove particularly valuable in directions 
where the FIR/Smm emission flux is too weak to perform FIR/Smm polarimetry.

\subsection{\label{Zeeman}Zeeman splitting measurements}

The line-of-sight magnetic field in dense, neutral (atomic or molecular) 
regions can in principle be measured directly through Zeeman splitting 
of radio spectral lines. 
In practice, though, near the GC, the task is made difficult 
by the very broad linewidths of GC clouds and by the line-of-sight 
blending of physically unrelated spectral features.
The first factor sets a stringent observational limit
($B_\parallel \sim {\rm a~few}\ 0.1$~mG), below which magnetic fields
cannot be detected with this method.

The first Zeeman splitting measurements toward the GC date back 
to the early 1990s and pertain to the CND.
\cite{schwarz&l_90} measured the Zeeman splitting 
of the H{\sc i} 21-cm absorption line
and reported the tentative detection of 
$B_\parallel \sim +0.5$~mG\footnote{\label{note}
The convention for the sign of $B_\parallel$ in the Zeeman splitting
community is opposite to that adopted in the Faraday rotation community.
Here, a positive (negative) value of $B_\parallel$ corresponds to 
a magnetic field pointing away from (toward) the observer.
Note that in their original paper, \cite{schwarz&l_90} mistakenly quoted 
a negative $B_\parallel$; the error was later corrected by
\cite{killeen&lc_92}.
} 
near the northern edge of the CND as well as a $3 \sigma$ upper limit 
$\sim 1.5$~mG near the southern edge.
From the Zeeman splitting of the OH 1667-MHz absorption line,
\cite{killeen&lc_92} derived $B_\parallel \sim -2$~mG both 
in the southern part (firm detection) and in the northern part 
(marginal detection) of the CND; for other OH clouds within 
the central $\sim 200$~pc, they obtained $3 \sigma$ upper limits 
$\sim (1-2)$~mG.
\cite{marshall&ly_95} and \cite{plante&lc_95} searched further for
Zeeman splitting in H{\sc i} 21-cm absorption over different areas
of the CND.
While the former found only an upper limit $\sim 0.5$~mG
for each of the northern and southern parts of the CND,
the latter reported 7 detections (1 positive and 6 negative values 
of $B_\parallel$) ranging between $-4.7$~mG and $+1.9$~mG
toward the northern part of the CND; the 6 strongest detections 
were argued to arise from the Northern Streamer rather than 
from the CND itself.

The disparate Zeeman results obtained for the CND are not necessarily
contradictory; they can be reconciled if $B_\parallel$ varies
substantially -- especially if $B_\parallel$ changes sign --
across the CND.
In that case, averaging over broad portions of the CND lowers 
the overall Zeeman signal, and sometimes does so to below 
the detectability threshold \citep{marshall&ly_95,plante&lc_95}.

Zeeman splitting measurements outside the CND were performed by
\cite{uchida&g_95}, who observed 13 selected positions within a few 
degrees of the GC, in the OH 1665-MHz and 1667-MHz absorption lines. 
The large velocities and broad linewidths of the absorption features
together with the relatively high molecular densities required 
to produce the observed absorption made the authors confident
that the absorbing clouds are located close to the GC.
All the measurements led to non-detections, with $3 \sigma$ upper limits 
to $B_\parallel$ $\sim (0.1-1)$~mG.
Here, beam dilution of the Zeeman signal, added to line-of-sight
averaging, could be partly responsible for the absence of detections.
Nonetheless, \cite{uchida&g_95}'s negative results provide evidence 
that the mean (or uniform) magnetic field through GC molecular clouds 
is either weaker than the mG field believed to thread NRFs 
or nearly perpendicular to the line of sight.
The latter possibility seems ruled out for some of the considered clouds,
notably, for the cloud associated with the Arched Filaments,
whose observed line-of-sight--velocity gradient should be accompanied
by field-line shearing along the line of sight.
More generally, FIR/Smm polarimetry indicates that 
the magnetic field inside molecular clouds is approximately horizontal
(see Sect.~\ref{IR_Smm}), which makes it unlikely that all the clouds
studied by \cite{uchida&g_95} would have their magnetic field nearly
perpendicular to the line of sight.
From this, one might conclude that the mean magnetic field through 
GC molecular clouds is generally below the mG level.

On the other hand, \cite{crutcher&rmt_96} mapped the Zeeman effect
in H{\sc i} 21-cm absorption toward the Main and North cores of Sgr~B2,
and they reported values of $B_\parallel$ at 4 different locations, 
all between $\simeq -0.1$~mG and $-0.8$~mG.
Sgr~B2 corresponds to one of the positions examined by \cite{uchida&g_95},
for which they managed to derive neither a value of $B_\parallel$ 
nor an upper limit to $B_\parallel$.
This failure to obtain information on $B_\parallel$ was due to
a combination of coarse angular resolution ($\sim 8'$, as opposed to 
$\sim 10''$ for \cite{crutcher&rmt_96}'s map), small-scale 
magnetic-field structure and blending of the OH absorption lines 
with OH maser emission lines.

Zeeman splitting of OH (1720~MHz) maser emission from the Sgr~A region
was measured by \cite{yusef&rgf_96, yusef&rgf_99}. 
The preliminary analysis of \cite{yusef&rgf_96} yielded very strong
magnetic fields, 
with $B_\parallel \sim +\,(2.0 - 3.7)$~mG in the Sgr~A East shell
and $B_\parallel \sim -\,(3.0 - 4.0)$~mG in the CND.
Follow-up, higher-resolution observations by \cite{yusef&rgf_99}
confirmed the presence of very strong fields, with $B_\parallel$ 
reaching $\sim +3.7$~mG in Sgr~A East and $\sim -4.8$~mG in the CND.
One should, however, keep in mind that OH masers arise in very special 
environments, for instance, in highly compressed regions behind
interstellar shock waves.
Therefore, field strengths inferred from OH maser emission lines 
may {\it a priori} not be considered typical of the ISM near the GC.

Like in Faraday rotation studies,
Zeeman splitting results have to be taken with caution.
When a true detection is made, it provides only a lower limit to 
the local $B_\parallel$ in the observed region, insofar as the Zeeman signal
is reduced (sometimes severely) by averaging over the observed area 
and along the line of sight.
In addition, $B_\parallel$ itself represents only a fraction of 
the total field strength, though not necessarily a small fraction
as in Faraday rotation studies, given that the magnetic field in dense
regions is probably nearly horizontal (see Sect.~\ref{IR_Smm}).
It should also be emphasized that the magnetic field in dense regions 
might not be representative of the magnetic field in the general ISM: 
like in the Galactic disk, molecular clouds near the GC are likely
to harbor much stronger fields than their surroundings.

\section{\label{discussion} Discussion}

A fragmentary picture of the interstellar magnetic field 
in the GC region (over $\sim 300$~pc along the Galactic plane 
and $\sim 150$~pc in the vertical direction)
emerges from the overview given in Sect.~\ref{observ}.
In the diffuse intercloud medium, sampled by the observed NRFs,
the field appears to be approximately poloidal on average, 
with considerable scatter (see Sect.~\ref{radio}), 
whereas in dense interstellar clouds, probed by FIR/Smm polarimetry,
the field appears to be approximately horizontal (see Sect.~\ref{IR_Smm}).
The field strength, $B$, is still a matter of controversy:
in NRFs, $B$ lies somewhere between $\sim 100~\mu$G
(equipartition/minimum-energy field strength, supported by inverse 
Compton scattering data) and $\gtrsim 1$~mG (dynamical condition 
$P_{\rm mag} \gtrsim P_{\rm ram}$);
in the general intercloud medium, $B$ lies between $\sim 10~\mu$G
(minimum-energy field strength) and $\sim 1$~mG (assumption of 
pressure balance with the NRFs);
and for dense interstellar clouds, some Zeeman splitting measurements
yield $|B_\parallel| \sim (0.1-1)$~mG (with values up to a few mG 
in the Sgr~A region), while others lead only to upper limits 
$\sim (0.1-1)$~mG.

\subsection{\label{field_direction}Magnetic field direction:
theoretical interpretation}

The approximately poloidal direction of the large-scale magnetic field 
in the intercloud medium near the GC can be explained in different ways.
The simplest explanation, originally proposed by \cite{sofue&f_87}, 
is that the inflow of Galactic disk matter that presumably created the CMZ 
dragged along the primordial Galactic magnetic field
and compressed its vertical component into the CMZ,
while letting its horizontal component diffuse away 
perpendicular to the disk.
This idea was elaborated upon by \cite{chandran&cm_00},
who emphasized the importance of vertical ambipolar diffusion 
in removing horizontal magnetic flux, especially in the CMZ
where the vertical ambipolar diffusion velocity (which is proportional
to $B^2$) reaches particularly high values.
Neglecting outward turbulent diffusion, \cite{chandran&cm_00} estimated 
that a pregalactic vertical field $\sim 0.2~\mu$G was required to account 
for the present-day CMZ field, which they supposed to be $\sim 1$~mG.

A pregalactic magnetic field as strong as $0.2~\mu$G might be unrealistic,
unless field amplification occurred during the very process of galaxy
formation -- for instance, through turbulent dynamo action in the
protogalaxy \citep{kulsrud&cor_97,kulsrud&z_08}.
But even if the pregalactic field was weaker,
the scenario of inward field advection might still be adequate
provided that the large-scale vertical field was rapidly amplified 
(to nearly its present-day value) by a dynamo operating in the Galactic 
disk and that its direction remained constant both in time and in space
(at least out to $\sim 10$~kpc from the GC).
To show this quantitatively, we now derive a rough upper limit to 
the CMZ field strength that might be expected under these conditions.
At the Galactic position of the Sun ($r_\odot = 8.5$~kpc),
the present-day large-scale vertical field can be inferred
from Faraday rotation measures of high-latitude Galactic pulsars 
and extragalactic radio sources; \cite{han&mq_99} thus derived 
$B_z \simeq 0.37~\mu$G, while Sui et al. (2009) recently obtained 
$B_z \simeq 0~\mu$G toward the North Galactic Pole and 
$B_z \simeq 0.5~\mu$G toward the South Galactic Pole.
Here, we adopt $B_z \simeq 0.25~\mu$G as a conservative estimate.
We further assume that the interstellar plasma, together with 
the frozen-in field lines, slowly drift inward,
with a radial velocity at the solar circle 
$v_r \simeq -(0.5 - 1.0)~{\rm km~s}^{-1}$
\citep[taken from the Galactic evolution models of][]{lacey&f_85}.
If these values apply over most of the $\sim 10^{10}$~yr lifetime 
of the Galaxy and if the inflowing vertical magnetic flux accumulates 
inside the CMZ at roughly the same rate as it crosses the solar circle, 
then the resulting present-day vertical field in the CMZ 
(whose projection onto the Galactic plane can be approximated as 
a $500~{\rm pc} \times 200~{\rm pc}$ ellipse; see Sect.~\ref{gas_distri}) 
is $\sim (0.85 - 1.7)$~mG.
This estimate is in surprisingly close agreement with the $\sim 1$~mG
observational value assumed by \cite{chandran&cm_00}.
However, the agreement may be a little fortuitous, 
and one has to bear in mind that our estimate
is contingent upon a number of important and uncertain hypotheses,
such as rapid dynamo amplification in the Galactic disk, 
no sign reversals in the large-scale vertical field, 
steady-state inflow and accumulation into the CMZ,
negligible turbulent diffusion out of the CMZ...
If one or more of these hypotheses fail, the actual CMZ field 
is less than estimated here. 

Another category of scenarios rely on outflows from the Galactic nucleus.
In one such scenario, a magnetic field was generated 
in the accretion disk around the central black hole, 
through creation of a seed field by a Biermann battery and subsequent 
amplification of this seed field either by the large-scale shear alone 
or by a standard dynamo; 
the generated field was then expelled into the surrounding ISM 
by Galactic winds or collimated outflows
\citep{chakrabarti&rv_94}.\footnote{
In a related model \citep{heyvaerts&np_88}, the central accretion disk 
generates magnetic loops which break off and expand away from their source.
Although this model provides a possible explanation for the NRF
phenomenon, it does not discuss the origin of the large-scale
magnetic field in the CG region.
}
{\it A priori}, the strong shear in the accretion disk must have caused 
the generated field to be nearly toroidal. 
Therefore, if the accretion disk was parallel to the Galactic plane
and if the field was expelled horizontally, this process alone could be 
a good candidate to explain a {\it horizontal} large-scale magnetic field 
in the GC region, but not a {\it poloidal} field.
This horizontal field would only be a few $\mu$G, unless, for some reason,
the expelled field remained confined to the CMZ.
In this case, the horizontal large-scale magnetic field in the CMZ
(approximated again as a $500~{\rm pc} \times 200~{\rm pc}$ ellipse)
would be $\sim (0.25 - 0.6)$~mG, as can be found by using the expression 
of the nuclear field provided by \cite{chakrabarti&rv_94},
together with a central black hole mass $\simeq 4 \times 10^6~M_\odot$
\citep{gillessen&eta_09,ghez&swl_08}.
In reality, magnetic activity in the immediate vicinity of the central 
black hole is certainly more complex than described above.
As it turns out, recent NIR polarimetric imaging observations of Sgr~A$^*$, 
in conjunction with earlier radio, infrared and X-ray data,
provide loose evidence that the accretion disk might indeed be 
nearly parallel to the Galactic plane 
\citep[and references therein]{trippe&pog_07,meyer&sed_07}.
However, they also suggest that the nuclear field switches back and forth 
between toroidal (as expected from strong shear) and poloidal 
(consistent with recurrent vanishing of the accretion disk).
And even if the nuclear field were strictly toroidal, it could,
after ejection, be twisted by the Coriolis force into a more 
poloidal field.

Still in the category of outflow scenarios,
a poloidal large-scale magnetic field in the GC region could find 
its roots in stellar activity very close to the GC 
\citep[e.g.,][]{sofue_84, chevalier_92} --
for instance, in a relatively recent nuclear starburst 
\citep[e.g.,][]{blandhawthorn&c_03}.
To start with, stellar fields are continuously injected into
the ISM via stellar winds and supernova explosions;
stellar fields alone cannot account for the present-day interstellar 
magnetic field, but one could conceive that a very early generation 
of them provided the seed field for a local dynamo (see below).
More relevant here, the energy release at the GC must have driven
the interstellar matter into expanding motions and created a vertically 
elongated shell structure, or wall, into which interstellar field lines 
were pushed back and compressed.
This mechanism was first invoked by \cite{sofue_84} to explain
the $\Omega$-shaped radio lobe detected by \cite{sofue&h_84},
and it was further studied by \cite{lesch&csw_89} and by \cite{chevalier_92}.
The result is that a pre-existing horizontal field could have turned 
into a seemingly poloidal configuration (as seen from the Sun), 
while a pre-existing vertical field could have been substantially enhanced 
at the position of the shell.

Finally, with the strong shear flows and the highly turbulent
motions existing near the GC, dynamo action there is almost inescapable.
\cite{chandran&cm_00} argued that dynamo amplification in the CMZ
is unlikely, on the grounds that the magnetic field there 
is way above equipartition with the turbulence.
However, their argument is based on a supposed mG field, 
which we now regard as a probable overestimate
(see Sect.~\ref{field_strength}).
If we assume that the CMZ is characterized by 
a turbulent velocity $v_{\rm turb} \sim 15~{\rm km~s}^{-1}$ 
(value adopted by \cite{yusef&m_87c,yusef&m_87b} for the cloud random
velocity), a space-averaged hydrogen density 
$\langle n_{\rm H} \rangle \sim 300~{\rm cm}^{-3}$ (see Paper~I)
and a total-to-hydrogen mass ratio of 1.453 (see Sect.~\ref{gas_distri}), 
we find that turbulent pressure corresponds to the magnetic pressure 
of 0.2~mG field.
In consequence, the saturated field strength expected from dynamo action
in the CMZ is $\sim 0.2$~mG, and any field $\lesssim 0.2$~mG
would be dynamically consistent with dynamo amplification in the CMZ
-- or, more generally, in a region encompassing the CMZ and the ionized
gas around it.

If the large-scale magnetic field in the GC region was indeed amplified 
by a local dynamo,
its final configuration must depend on the shape of the dynamo domain.
\cite{moss&s_08} showed that in astrophysical objects composed
of a flattened disk-like structure and a quasi-spherical halo-like
structure (such as spiral galaxies), a dynamo generates a quadrupolar
(i.e., symmetric in $z$) field if the disk-like structure is more dynamo-active
and a dipolar (i.e., anti-symmetric in $z$) field if the halo-like structure 
is more dynamo-active.
This theoretical result applied to the GC region implies that
a dynamo field there must be quadrupolar if the disk-like CMZ dominates 
dynamo action and dipolar if the surrounding halo-like volume of ionized gas 
dominates.
The latter possibility seems {\it a priori} more likely, given that
the halo of ionized gas represents most of the volume, encloses most of
the interstellar mass and is highly turbulent, but one cannot draw any
definite conclusions without resorting to detailed numerical dynamo
calculations.
It is also noteworthy that, if the magnetic field in interstellar clouds 
is largely decoupled from that in the intercloud medium 
\citep[as suggested by][]{morris&s_96}, a local dynamo should naturally 
produce a dipolar field in the intercloud medium.
Either way, a poloidal (and hence dipolar) large-scale magnetic field 
in the GC region could very well be explained by a local dynamo.

The different scenarios described above refer to the large-scale
component of the interstellar magnetic field in the intercloud medium
near the GC.
We emphasize that they are not mutually exclusive.
In particular, it is quite possible (even likely, in our view) 
that inward advection from the Galactic disk, outflows from the Galactic 
nucleus and local dynamo amplification all came into play.
We also note that the wide scatter observed in the orientations 
of NRFs, especially the smaller ones, is almost certainly 
due to interstellar turbulence, regardless of the exact mechanism 
responsible for the appearance of filaments.

We now turn to the magnetic field inside GC molecular clouds.
According to \cite{chuss&ddd_03}, the reason why the field within dense clouds 
is approximately horizontal can be understood in two alternative ways.
A first possibility would be that the GC region was originally pervaded
by a large-scale azimuthal magnetic field, as could be the case 
if a strong toroidal field was generated in the accretion disk 
around the central black hole and then expelled horizontally 
into the surrounding ISM or if the field was amplified by a local 
dynamo dominated by the disk-like CMZ (see above).
In this scenario, the original field direction would have been preserved 
only in the dense and massive molecular clouds;
outside of them, it would have been distorted by winds due to supernovae.
However, for the reasons outlined above, it is more likely that
the GC region was originally pervaded by a large-scale poloidal
magnetic field. 
In molecular clouds, this field would have been sheared out 
in a horizontal direction, either by the cloud bulk motions
(from differential rotation or from turbulence) with respect to 
the diffuse intercloud medium \citep{novak&crg_03,chuss&ddd_03}
or by the forces (of compressive or tidal nature) that created 
and/or shaped the clouds \citep[and references therein]{morris&s_96}.

Along different lines, \cite{morris&s_96} suggested that
the magnetic field within molecular clouds is only loosely coupled 
to the intercloud field \citep[see also][]{morris_07}.
Decoupling between cloud and intercloud fields is easily achieved
if, as expected, clouds are rotating \citep[e.g.,][]{fletcher&ks_09}.
Initially, clouds must be magnetically connected to the diffuse
intercloud medium from which they form.
However, as they contract and spin up, they rapidly wind up the field 
lines that thread them, until, following magnetic reconnection 
or ambipolar diffusion, the wound-up field inside them becomes detached 
from the external field.
Since clouds generally rotate about roughly vertical axes, their final
internal fields will be predominantly horizontal.

\subsection{\label{field_strength}Magnetic field strength:
critical discussion}

At the present time, the most uncertain aspect of the large-scale
interstellar magnetic field in the GC region is its strength, 
with current estimates ranging from $\sim 10~\mu$G to $\gtrsim 1$~mG 
in the diffuse ISM.
The low values are observational estimates 
obtained for a supposed equipartition/minimum-energy state,
while the high values are theoretical estimates 
based on pressure balance considerations.
Both types of estimates involve some questionable assumptions and 
raise a number of problems, which we now briefly review and discuss,
starting with the high-$B$ estimates.

The reasoning leading up to the picture of a pervasive mG magnetic field
hinges on two separate arguments (see beginning of Sect.~\ref{observ}).
First, the apparent resistance of NRFs to collisions with ambient
molecular clouds implies $P_{\rm mag} \gtrsim P_{\rm ram}$, 
which translates into $B \gtrsim 1$~mG, inside NRFs.
Second, pressure confinement of NRFs requires $B \gtrsim 1$~mG 
in the external ISM as well.
Let us examine both arguments in turn.

The first argument already presents some weaknesses.
It is true that the majority of NRFs are nearly straight,
even though a few of them appear to be significantly distorted
and a few others could potentially have deformations that escape 
detection from Earth because of projection effects.
A moot point, however, is whether NRFs are truly colliding 
with molecular clouds.
\cite{morris&s_96} pointed out that all the well-studied NRFs 
at the time of their writing showed definite signs 
(e.g., slight bending, brightness discontinuity) 
of physical interactions with adjacent molecular clouds.
On the other hand, \cite{yusef_03} noted that dynamical collisions
with molecular clouds should produce OH (1720~MHz) maser emission,
yet none had been detected at the apparent interaction sites.
In addition, some NRFs do not seem to have any molecular cloud 
associated with them \citep[e.g.,][]{lang&akl_99,lang&me_99}.

Another issue concerns the exact constraint imposed by the rigidity
of a truly colliding NRF.
\cite{yusef&m_87b} derived the condition $P_{\rm mag} \gtrsim P_{\rm ram}$,
and they adopted $n_{\rm H_2} \sim 10^4~{\rm cm}^{-3}$ to estimate 
$P_{\rm ram}$.
This value of $n_{\rm H_2}$ is probably too high, as the filling factor 
of molecular clouds with $n_{\rm H_2} \gtrsim 10^4~{\rm cm}^{-3}$ 
is only $\lesssim 3\%$ (see Paper~I) 
or even as low as $\sim 1\%$ \citep{goto&ung_08},
so that the likelihood that a given NRF 
be colliding with such a high-density cloud is rather low.
Most of the NRFs undergoing a cloud collision could actually be colliding 
with a lower-density cloud or with the low-density envelope of a dense cloud 
(Boldyrev, private communication).
More importantly, the condition $P_{\rm mag} \gtrsim P_{\rm ram}$
itself may in fact be too stringent.
\cite{chandran_01} showed that a filament interacting with a cloud 
at a single location along its length has only small distortions
so long as the Alfv\'en speed in the intercloud medium, 
$V_{\rm A} = B / \sqrt{4 \pi \rho}$, is much greater than 
the cloud velocity, $v_{\rm cloud}$.
With $\rho = 1.453 \, m_{\rm p} \, n_{\rm H}$,
$\langle n_{\rm H} \rangle \sim (1-10)~{\rm cm}^{-3}$
in the intercloud medium (see Paper~I)
and $v_{\rm cloud} \sim 15~{\rm km~s}^{-1}$
\citep[as in][]{yusef&m_87c,yusef&m_87b},
the easier condition $V_{\rm A} \gg v_{\rm cloud}$ is equivalent to 
$B \gg (8-26)~\mu$G.

The bottom line is that
for some NRFs, like those of the Radio Arc, which remain nearly straight 
despite multiple physical interactions with molecular clouds 
\citep{morris_90}, the original conclusion that $B \gtrsim 1$~mG 
seems reasonably solid.
For other NRFs, the evidence for a mG field is less compelling,
or even completely absent.

The second argument is even more questionable.
\cite{morris_90} reasoned that NRFs must be pressure-confined
and, considering the possibility of plasma confinement,
he estimated the thermal pressure of the very hot gas
at $n \, T \sim 10^7~{\rm cm}^{-3}~{\rm K}$.
Since this is much less than the pressure of a 1~mG magnetic field, 
$P_{\rm mag} / k \simeq 3 \times 10^8~{\rm cm}^{-3}~{\rm K}$,
he concluded that the confining pressure cannot be thermal, 
but instead must be magnetic.
This line of reasoning raises several important issues, 
some of which might possibly invalidate it. 

First, \cite{morris_90} might have considerably underestimated
the very hot gas thermal pressure.
\cite{koyama&mst_96} presented X-ray spectroscopic imaging observations
of a $1^\circ \times 1^\circ$ area around the GC, which 
point to a much higher value.
Attributing the diffuse X-ray continuum emission to thermal bremsstrahlung, 
they derived an electron temperature $k T_{\rm e} \gtrsim 10$~keV 
($T_{\rm e} \gtrsim 10^8$~K) and an electron density 
$n_{\rm e} \sim (0.3 - 0.4)~{\rm cm}^{-3}$.
If ions and electrons are in collisional equilibrium, 
the corresponding total thermal pressure is 
$n \, T \gtrsim (6-8) \times 10^7~{\rm cm}^{-3}~{\rm K}$.
But \cite{koyama&mst_96} found evidence that the thermal pressure 
of the ions might be at least one order of magnitude higher than that 
of the electrons.
If this were to be confirmed, the very hot gas would be able to supply 
the required confining pressure for NRFs with mG fields.
Interestingly, the $\simeq 270$~pc radial extent of the very hot gas region 
\citep{yamauchi&kkk_90} would then naturally explain why NRFs are observed 
over a $\sim 300$~pc region \citep{larosa&nlk_04} only.
However, we have to emphasize that the plasma pressure inferred from
\cite{koyama&mst_96}'s work might be a little extreme. 
More recently, \cite{muno&bbf_04} observed a smaller,
$17' \times 17'$ area around the GC, with better signal-to-noise ratio 
and finer angular resolution. 
To explain the observed diffuse X-ray emission, they had to appeal to 
two plasma components: a soft component, with $k T \simeq 0.8$~keV and 
$n_{\rm e} \sim (0.1 - 0.5~{\rm cm}^{-3}) \, (d / 50~{\rm pc})^{-1/2}$,
and a hard component, with $k T \simeq 8$~keV and
$n_{\rm e} \sim (0.1 - 0.2~{\rm cm}^{-3}) \, (d / 50~{\rm pc})^{-1/2}$,
where $d$ is the line-of-sight depth of the X-ray emitting region.
Both components were assumed to be in collisional-ionization equilibrium,
consistent with the measured energies and flux ratios of the spectral lines.
Under these conditions and with $d \simeq 270$~pc \citep{yamauchi&kkk_90}, 
the total thermal pressure of the plasma is 
$\sim (1-2) \times 10^7~{\rm cm}^{-3}~{\rm K}$,
close to \cite{morris_90}'s original estimate and too low to confine NRFs.
To be sure, the above pressure estimates make sense only to the extent that 
the observed diffuse X-ray emission is truly produced by a thermal plasma.
These estimates do not allow us to conclude either way on the potential
ability of hot plasma to confine NRFs.

Another interesting possibility is that NRFs could be confined by
magnetic tension forces \citep{lesch&r_92,uchida&g_95}.
\cite{yusef&m_87b} noticed that the linear filaments of the Radio Arc
appear to be surrounded by a helical magnetic structure winding about them.
Having estimated that magnetic pressure in the filaments is much higher
than the ambient gas pressure, they advanced the view that 
the magnetic field is in a local force-free state.
Accordingly, the electric current flows along field lines,
and the field-aligned current induces a locally toroidal field component, 
in agreement with the observed helical structure. 
This toroidal component, in turn, provides a confining magnetic tension force.
Observational evidence for helical or twisted magnetic fields
has been found in filaments other than those of the Radio Arc
\citep[e.g.,][]{gray&nec_95,yusef&wp_97,morris&ud_06}.
From a theoretical point of view, field-aligned currents and helical fields 
can be explained in various ways.
\cite{benford_88} proposed an electrodynamic model for the Radio Arc,
wherein the motion of a partially ionized molecular cloud across 
a strong poloidal magnetic field drives electric currents around
a closed-loop circuit 
originating at the cloud and continuing along the poloidal field lines 
\citep[see also][]{morris&y_89}.
For the Snake, \cite{bicknell&l_01} envisioned the picture of a magnetic 
flux tube having both ends anchored in rotating molecular clouds;
in this picture, the differential rotation between both clouds generates 
a locally toroidal field component and, hence, give rise to a helical pattern.
Each of these two models could potentially be applied to other NRFs,
bearing in mind that twisted magnetic fields are subject to the kink 
instability.
In any event, even if the operation of magnetic tension forces could be 
firmly established, it would still be difficult to quantify their actual
contribution to the confinement of NRFs without a detailed knowledge 
of the exact magnetic configuration in and around NRFs.

Finally, and most importantly, NRFs do no need to be confined at all;
they could very well be transient or dynamic structures 
out of mechanical balance with their surroundings.
As a matter of fact, several plausible models for their origin
describe them in these non-equilibrium terms.
In the cometary-tail model of \cite{shore&l_99},\footnote{
Neither the cometary-tail model of \cite{shore&l_99}
nor the turbulent model of \cite{boldyrev&y_06} is able to explain 
the Radio Arc, which differs from the other isolated NRFs in that 
it is clearly organized on a large scale.
}
NRFs are the long and thin magnetic wakes produced 
by a weakly magnetized Galactic wind impinging on GC molecular clouds.
The advected magnetic field drapes around the clouds and stretches out 
behind them, growing to the point where its pressure balances 
the ram pressure of the Galactic wind.
For typical wind parameters, the field inside the wakes thus reaches
$\sim 1$~mG, independent of the ambient field strength, which could be
as low as $\sim 10~\mu$G (estimate based on the predicted wake lengths
and on stability considerations).
The overall orientation of the wakes is governed by the direction 
of the wind, which, the authors claim, is roughly vertical.
Hence, the observed tendency of NRFs to run vertical would reflect
the direction of the wind rather than that of the large-scale magnetic 
field (as usually presumed).
An advantage of this dynamical model is that it avoids the MHD
stability problems faced by static equilibrium models.
It also provides an efficient mechanism to accelerate electrons
to relativistic energies, through wave-particle interactions 
in the turbulent cascades driven near the current sheets of the wakes.

Still in the spirit of a dynamic ISM, NRFs could be regarded as
direct manifestations of the intense turbulent activity characterizing
the GC region \citep{boldyrev&y_06}.
Turbulence there naturally leads to a highly intermittent magnetic field
distribution, with strongly magnetized filamentary structures
arising in an otherwise weak-field (e.g., $B \sim 10~\mu$G) background.
Because the turbulent intensity varies on a scale comparable to or only
slightly larger than the outer scale of the turbulence, magnetic flux
is expelled from the region of intense turbulence (diamagnetic pumping)
before the field has a chance to grow strong everywhere.
In this context, the observed NRFs would be nothing else than the strongly 
magnetized filaments, and their field strength, set by the external
turbulent pressure, would typically be $\gtrsim 0.1$~mG.
The preferentially vertical direction of the longest NRFs could be
related to the $\Omega$-shaped radio lobe detected by \cite{sofue&h_84}; 
more generally, it could be a direct consequence of the roughly poloidal 
geometry of the large-scale magnetic field. 
As for the synchrotron-emitting electrons, they could be either
pervasive or concentrated near localized sources.

Aside from the possible shortcomings in the derivation of a mG field,
the claims for such a strong field have elicited a number of important
criticisms.
A frequently voiced objection is that the ensuing synchrotron lifetimes, 
$t_{\rm syn}$, are too short.
The synchrotron lifetime is indeed a strongly decreasing function of
field strength, given by $t_{\rm syn} \propto E^{-1} \, B_\perp^{-2}$
for a relativistic electron of energy $E$
and by $t_{\rm syn} \propto \nu^{-1/2} \, B_\perp^{-3/2}$
at the corresponding synchrotron frequency, $\nu \propto B_\perp \, E^2$.
At the 74~MHz and 330~MHz frequencies of the diffuse nonthermal emission
detected by \cite{larosa&bsl_05}, the synchrotron lifetimes
in a $B_\perp = 1$~mG field would only be 
$t_{\rm syn} \simeq 1.2 \times 10^{5}$~yr and 
$t_{\rm syn} \simeq 0.6 \times 10^{5}$~yr, respectively.
Unless the source of nonthermal emission is short-lived, relativistic
electrons would need to be injected or re-accelerated on such short
timescales. \cite{larosa&bsl_05} considered this to be implausible,
whereas \cite{morris_07} estimated that the local supernova rate 
is more than sufficient to meet the required timescales.

At the higher 1.5~GHz and 5~GHz frequencies of many NRF observations,
the synchrotron lifetimes in a 1~mG field would be even shorter:
$t_{\rm syn} \simeq 2.7 \times 10^{4}$~yr and 
$t_{\rm syn} \simeq 1.5 \times 10^{4}$~yr, respectively
(supposing $B_\perp \simeq B$, a reasonable assumption in the case of NRFs).
Here, the trouble is not so much that these short lifetimes would impose 
similarly short injection/re-acceleration timescales --
as explained above, NRFs could be short-lived structures, 
and if they are not, one could easily imagine that they are fueled 
by local, long-lived sources of relativistic electrons.
The real concern is that electrons might not be able to travel far enough
\citep[e.g.,][]{yusef_03, morris_07}.
Presumably, relativistic electrons stream along field lines 
at about the Alfv\'en speed in the ionized gas, 
$V_{\rm A,i} = B / \sqrt{4 \pi \rho_i}$,
such that, over their lifetimes $t_{\rm syn}$, they travel distances
$d \simeq V_{\rm A,i} \, t_{\rm syn} 
\propto \nu^{-1/2} \, \rho_i^{-1/2} \, B^{-1/2}$.
The ionized gas density, $\rho_i$, is quite uncertain, as witnessed by
the widely different values adopted by different authors.
Since the ionized phase of the ISM is largely dominated by its warm 
component, where helium is only weakly ionized, we may let
$\rho_i \simeq m_{\rm p} \, n_{\rm e}$. 
If we further approximate the free-electron density, $n_{\rm e}$, 
by its space-averaged value from \cite{cordes&l_02}'s model,
we find that $n_{\rm e} \simeq 10~{\rm cm}^{-3}$ close to the GC 
and $n_{\rm e} \gtrsim 1~{\rm cm}^{-3}$ out to $\simeq 220$~pc 
along the Galactic plane and up to $\simeq 40$~pc along the vertical,
i.e., in most of the region of interest.
For a 1~mG field, the Alfv\'en speed is $V_{\rm A,i} \simeq 
(2200~{\rm km~s}^{-1}) \, (n_{\rm e} / 1~{\rm cm}^{-3})^{-1/2}$ 
and the traveled distances at 1.5~GHz and 5~GHz are
$d \simeq (60~{\rm pc}) \, (n_{\rm e} / 1~{\rm cm}^{-3})^{-1/2}$ and 
$d \simeq (33~{\rm pc}) \, (n_{\rm e} / 1~{\rm cm}^{-3})^{-1/2}$, 
respectively.
These calculated distances are compatible with the observed lengths 
of most NRFs, but they are difficult to reconcile with some NRFs being 
as long as $\sim 60$~pc.
This difficulty can be circumvented 
if the longest NRFs lie in a particularly low-$n_{\rm e}$ environment, 
or if particle injection occurs at more than one location along them 
\citep[for instance, the Snake, which is $\sim 60$~pc long, has two kinks 
along its length, one or both of which could conceivably be associated 
with particle injection; e.g.,][]{bicknell&l_01}, 
or else if re-acceleration takes place more-or-less continuously 
along them \citep[e.g.,][]{morris_07}.
Incidentally, re-acceleration also provides a possible explanation 
for the observed constancy of the radio spectral index along the lengths
of several NRFs \citep[e.g.,][]{lang&me_99, larosa&klh_00}, although
it is not immediately clear why re-acceleration would precisely
counteract the spectral steepening due to synchrotron losses.

The above discrepancy between the calculated distances traveled by
relativistic electrons in a 1~mG field and the observed lengths
of the longest NRFs is only marginal. 
Furthermore, there exist viable ways, other than decreasing 
the field strength, to bring them into agreement. 
But more fundamentally, the discrepancy poses a problem only 
to the extent that NRFs are ``illuminated flux tubes'', i.e., 
flux tubes into which relativistic electrons have been injected.
Should they instead be regions of compressed magnetic field
\citep[e.g.,][]{shore&l_99,boldyrev&y_06}, the mismatch 
would become totally irrelevant.
For all these reasons, we feel that the short synchrotron lifetimes
and the associated short distances traveled by relativistic electrons
may not be held up against 
a mG field inside NRFs.

To conclude our discussion of the high-$B$ estimates, 
the claim that NRFs have a mG field may have to be toned down.
While there is good evidence that a fraction of them do,
others could possibly have a weaker field.
As for the general ISM, we see no cogent reason to believe that 
it is magnetized at the mG level.

We now turn to the low-$B$ estimates and discuss the validity of 
the equipartition/minimum-energy assumption, which, for brevity,
we will refer to as the equipartition assumption.
This hypothesis admits no rigorous theoretical justification,
yet it seems to yield acceptable results in large-scale regions 
of the Galactic disk as well as in external galaxies as wholes
\citep[e.g.,][]{beck_01,beck&k_05}.
In these objects, the large-scale magnetic field is predominantly
horizontal, the total field strength is typically a few $\mu$G
and cosmic-ray sources (e.g., supernova shocks) likely abound.
As a result, vast amounts of cosmic rays are injected into the ISM, 
where the horizontal field lines tend to keep them confined.
Cosmic-ray pressure can then build up until it reaches a value, 
presumably $\sim P_{\rm mag}$, that is high enough to break 
the magnetic confinement -- for instance, via the Parker instability.
Not only does this instability enable cosmic rays to escape from 
the galactic disk through the formation and rupture of giant magnetic loops,
but it also leads to magnetic field amplification through 
enhanced dynamo action \citep{parker_92,hanasz&kol_04,hanasz&olk_09}.
Both effects conspire to maintain cosmic-ray pressure comparable
to magnetic pressure.

The situation near the GC is completely different. 
Cosmic-ray sources are undoubtedly much more abundant there 
than in the Galactic disk.
On the other hand, the large-scale magnetic field is approximately 
vertical, so that cosmic rays streaming along field lines naturally 
follow the shortest path out of the Galaxy \citep{chandran&cm_00}.
In addition, if the field is as strong as $\sim 1$~mG, cosmic rays 
stream away at a very high speed ($\sim V_{\rm A,i}$).\footnote{
Regardless of the field strength, it is likely that the powerful winds 
emanating from the GC \citep[e.g.,][]{blandhawthorn&c_03} or produced by 
supernova explosions also contribute to cosmic-ray escape.
However, since these winds tend to evacuate magnetic fields as well,
it is not clear that their net effect is to reduce cosmic-ray pressure
relative to magnetic pressure.
}
Thus, both the injection and the escape of cosmic rays proceed 
at a much faster rate than in the disk. 
More to the point, the two processes are not related by a self-regulating 
mechanism, such as the ``cosmic-ray valve'' operating in the disk, 
which would, at the same time, keep cosmic-ray and magnetic pressures 
close to equipartition.

Observationally, the obvious rigidity and organized structure of NRFs 
strongly support the idea that they are magnetically dominated 
\citep[e.g.,][]{anantharamaiah&peg_91,lang&me_99}.
The case of the general ISM is not so self-evident.
Observations of the diffuse $\gamma$-ray emission from the GC region 
reveal a pronounced $\gamma$-ray excess within $\sim 0\fdg6$ of the GC
\citep{mayer&bde_98}, which could speak in favor of an elevated 
cosmic-ray pressure there.
However, it is unlikely that this $\gamma$-ray excess can entirely
be attributed to cosmic-ray interactions with interstellar matter. 
In fact, \cite{mayer&bde_98} showed that their $\gamma$-ray data
could be reproduced with a combination of unresolved compact sources 
(such as pulsars) and a truly diffuse interstellar contribution 
from cosmic rays having the same density as in the inner Galaxy.
This result, combined with \cite{larosa&bsl_05}'s study 
of the diffuse nonthermal radio emission from the GC region,
lends some credence to the equipartition assumption.

To conclude our discussion of the low-$B$ estimates, 
NRFs probably have super-equipartition magnetic fields,
whereas the general ISM might have its magnetic field in rough 
equipartition with cosmic rays.

New observations with the Fermi gamma-ray space telescope 
will hopefully shed more light on this topic,
although using the Fermi data to obtain better constraints 
on the cosmic-ray electron energy density near the GC might prove 
rather tricky.
In principle, the task will require
(1) removing the foreground $\gamma$-ray emission along the long line 
of sight toward the GC region,
(2) isolating the diffuse interstellar emission produced by cosmic rays,
(3) separating the contributions from cosmic-ray nuclei ($\pi^0$ decay)
and electrons (bremsstrahlung and inverse Compton scattering)
and (4) having a good knowledge of both the interstellar matter
distribution (for the $\pi^0$ decay and bremsstrahlung components) 
and the interstellar radiation field (for the inverse Compton component).
One difficulty for cosmic-ray electrons is that their contribution
to $\gamma$-ray emission in the energy range covered by Fermi 
(ideally $\sim 30~{\rm MeV} - 300~{\rm GeV}$)
is overshadowed by the contribution from cosmic-ray nuclei.
On the other hand, the inverse Compton component, observed slighty
above or below the Galactic plane, is expected to be less contaminated
by foreground emission, insofar as the photon density is much higher 
in the Galactic bulge than along the line of sight.
Altogether, there is good hope that Fermi observations of the inverse 
Compton emission from the GC region will help to determine the cosmic-ray 
electron density with better accuracy, and, when coupled with 
the low-frequency radio continuum data, help to constrain 
the magnetic field strength near the GC.
 
Faraday rotation studies do not add any significant constraints 
on the interstellar magnetic field strength.
The inferred values of $|B_\parallel|$ are typically a few $\mu$G, 
with considerable uncertainty (see Sect.~\ref{Faraday}).
These values of $|B_\parallel|$ indicate that the large-scale magnetic field 
in the diffuse ionized medium is $\gtrsim$ a few $\mu$G,
which is compatible with both an equipartition field $\sim 10~\mu$G
and a dynamically dominant field $\sim 1$~mG.
However, if the large-scale field has a roughly poloidal geometry, 
with only a small component along the line of sight, the equipartition 
estimate might be a little too low to explain the Faraday rotation results.

For dense neutral clouds, Zeeman splitting studies lead to a mixture 
of positive detections, with $|B_\parallel| \sim (0.1-1)$~mG 
(outside the Sgr~A region), and non-detections, 
with $|B_\parallel| \lesssim (0.1-1)$~mG, 
again subject to important uncertainty as well as possible dilution 
of the Zeeman signal (see Sect.~\ref{Zeeman}).
These mixed results can be understood if the magnetic field inside 
dense clouds is roughly horizontal, such that its line-of-sight component 
may lie anywhere between 0 and the total field strength.
The latter could then be $\sim 1$~mG or loosely range from a few 0.1~mG 
to a few mG.
A less direct, and even cruder, estimation of the field strength 
inside molecular clouds was made by \cite{chuss&ddd_03}, based on 
their Smm polarimetric observations. 
Interpreting the measured dependence of field direction on gas density 
in terms of shearing of an initially poloidal field, they came up with 
a characteristic field strength $\sim 3$~mG (see Sect.~\ref{IR_Smm}).

\section{\label{additional}Additional input}

\subsection{\label{connection}Connection with 
the rest of the Galaxy}

How does our view of the interstellar magnetic field in the GC region
fit in with what we know about the magnetic field in the Galaxy at large?
Are both magnetic systems connected in any way or are they completely
independent?

Our current knowledge of the overall distribution and morphology
of the Galactic magnetic field away from the GC relies primarily on
synchrotron emission and Faraday rotation studies.

Synchrotron intensity measurements give access to the total field strength 
distribution, subject to the assumption of equipartition between magnetic 
fields and cosmic rays (discussed in Sect.~\ref{field_strength}).
Based on the synchrotron map of \cite{beuermann&kb_85}, \cite{ferriere_98} 
thus found that the total field has a value $\simeq 5~\mu$G near the Sun,
a radial scale length $\simeq 12$~kpc and a local vertical scale height 
$\simeq 4.5$~kpc.
In addition, synchrotron polarimetry indicates that the local ratio of
ordered (regular + anisotropic random) to total fields is $\simeq 0.6$
\citep{beck_01}, implying an ordered field $\simeq 3~\mu$G near the Sun.

Faraday rotation measures of Galactic pulsars and extragalactic radio
sources also provide valuable information,
more specifically relevant to the uniform (or regular) magnetic field,
${\bf B}_{\rm u}$, in the ionized ISM. 
Here is a summary of what we have learnt from them, 
first on the strength and second on the direction of ${\bf B}_{\rm u}$.
From pulsar RMs, we now know that $B_{\rm u} \simeq 1.5~\mu$G near the Sun
\citep{rand&k_89,han&mlq_06} and that $B_{\rm u}$ increases toward the GC, 
to $\gtrsim 3~\mu$G at $r = 3$~kpc \citep{han&mlq_06}, 
i.e., with an exponential scale length $\lesssim 7.2$~kpc.
$B_{\rm u}$ also decreases away from the Galactic plane, 
albeit at a very uncertain rate -- extragalactic-source RMs suggest
an exponential scale height $\sim 1.4$~kpc \citep{inoue&t_81}.

In the Galactic disk, ${\bf B}_{\rm u}$ is nearly horizontal
and generally dominated by its azimuthal component.
Near the Sun, ${\bf B}_{\rm u}$ points toward $l \simeq 82^\circ$,
corresponding to a magnetic pitch angle $p \simeq -8^\circ$ 
\citep{han&mq_99} and implying a clockwise direction
about the $z$-axis (see Fig.~\ref{fig:coordinates}).
However, ${\bf B}_{\rm u}$ reverses several times with decreasing radius,
the number and radial locations of the reversals being still highly 
controversial \citep{rand&l_94, han&mq_99, vallee_05, han&mlq_06, 
brown&hgt_07, noutsos&jkk_08}.
These reversals have often been interpreted as evidence that
the uniform field is bisymmetric (azimuthal wavenumber $m \!=\! 1$),
although an axisymmetric ($m \!=\! 0$) field would be expected 
from dynamo theory.
Recently, \cite{men&fh_08} showed that neither the axisymmetric nor
the bisymmetric picture is consistent with the existing pulsar RMs,
and they concluded that the uniform field must have a more complex pattern.
Along the vertical, ${\bf B}_{\rm u}$ is roughly symmetric in $z$,\footnote{
A magnetic field is said to be symmetric (antisymmetric) in $z$,
or, equivalently, quadrupolar (dipolar), if its horizontal component 
is an even (odd) function of $z$ and its vertical component an odd (even) 
function of $z$.
}
at least close enough to the midplane \citep{rand&l_94}.

In the Galactic halo, ${\bf B}_{\rm u}$ could have a significant
vertical component, with $(B_{\rm u})_z \sim 0.37~\mu$G \citep{han&mq_99}
or $(B_{\rm u})_z \sim 0.25~\mu$G on average between the North and South 
Poles (Sui et al. 2009) at the position of the Sun. 
In contrast to the situation in the disk, the azimuthal component of 
${\bf B}_{\rm u}$ shows no sign of reversal with decreasing radius.
Along the vertical, ${\bf B}_{\rm u}$ is roughly antisymmetric in $z$ 
(counterclockwise at $z>0$ and clockwise at $z<0$) inside the solar circle 
and roughly symmetric (clockwise at all $z$) outside the solar circle
\citep{han&mbb_97, han&mq_99}.

\cite{sun&rwe_08} developed comprehensive 3D models of the Galactic
magnetic field, constrained by an all-sky map of extragalactic-source RMs
together with observations of the Galactic total and polarized emission 
over a wide range of radio frequencies.
They obtained a good fit to the data for axisymmetric models 
where the disk field is purely horizontal, has constant pitch angle 
$p = -12^\circ$, reverses inside the solar circle
and is symmetric in $z$ (clockwise near the Sun),
while the halo field is purely azimuthal and antisymmetric in $z$
(counterclockwise/clockwise above/below the midplane at all radii).
They also came to the conclusion that bisymmetric models are incompatible 
with the RM data.

Finding ${\bf B}_{\rm u}$ to have quadrupolar parity in the disk 
and dipolar parity in the halo is consistent with the predictions 
of dynamo theory and with the results of galactic dynamo calculations, 
even if, in the very long run (i.e., on timescales longer than current 
galactic ages), the field may eventually evolve toward a single-parity 
state \citep{brandenburg&dms_92, ferriere&s_00, moss&s_08}.
Moreover, the signs of $(B_{\rm u})_r/(B_{\rm u})_\theta$ in the disk
(negative throughout) and $(B_{\rm u})_z/(B_{\rm u})_\theta$ in the halo 
(negative/positive above/below the midplane) are those expected from 
azimuthal shearing of radial and vertical fields, respectively,
by the large-scale differential rotation of the Galaxy.

Let us now look into the possible connections between 
the Galactic magnetic field described above 
and the GC field discussed in the previous sections.
Here, we are only interested in the uniform component of the field,
which, for brevity, we will simply refer to as the field.
As explained at the beginning of Sect.~\ref{observ}, the observed
orientations of NRFs suggest that the GC field is approximately poloidal, 
and hence dipolar, in the diffuse intercloud medium.
However, they do not tell us whether the field is pointing north or south.

Faraday rotation studies could, under certain conditions, 
provide the sought-after sign information,
but unfortunately they lead to conflicting conclusions.
\cite{novak&crg_03} noted that the available RMs toward NRFs within $1^\circ$ 
of the GC exhibit a sign pattern indicative of an antisymmetric field 
running counterclockwise above the midplane and clockwise below it
(see Sect.~\ref{Faraday}) -- exactly as in the Galactic halo.
They further suggested that this pattern could result from
azimuthal shearing by the Galactic differential rotation 
of an initially vertical field pointing north ($B_z > 0$)
-- again as in the halo.
In contrast, \cite{roy&rs_08} found that the RMs of background
extragalactic sources seen through the area $(|l|<6^\circ,|b|<2^\circ)$
are mostly positive in each of the four quadrants, consistent with
a symmetric field pointing toward us.
Such a symmetric field cannot be produced by large-scale shearing,
or by any symmetric distortion, of a poloidal field.
It is, therefore, probably unrelated to the dominant poloidal field
and may not be used to constrain its (north or south) direction.

Zeeman splitting studies are not of great help here.
The only true detections outside the Sgr~A region and away from OH masers 
pertain to Sgr~B2, where \cite{crutcher&rmt_96} measured $B_\parallel < 0$
(see Sect.~\ref{Zeeman} and footnote~\ref{note}).
If the line-of-sight field in the dense clouds sampled by Zeeman
measurements has the same sign as in the surrounding intercloud medium, 
\cite{crutcher&rmt_96}'s results suggest ${\rm RM} > 0$ toward Sgr~B2.
Since Sgr~B2 lies in the $(l>0,b>0)$ quadrant, this RM sign
is in agreement with the general RM patterns of both \cite{novak&crg_03} 
and \cite{roy&rs_08}.

Irrespective of the exact RM pattern and of the sign of $B_z$ near the GC,
the predominantly poloidal, and hence dipolar, GC field could naturally
connect with the dipolar halo field.
Both fields together could actually form a single magnetic system, 
which, in turn, could be the outcome of a large-scale quasi-spherical 
dynamo -- corresponding, for instance, to an A0 (antisymmetric \& 
$m \!=\! 0$) dynamo mode \citep{han&mbb_97}.
However, one may not jump to the conclusion that the poloidal field
is a pure dipole, as proposed by \cite{han_02} and often assumed 
in the cosmic-ray propagation community
\citep[e.g.,][]{alvarez&es_02,prouza&s_03}.
A pure dipole can be expected in a current-free medium,
but not inside the highly conducting ISM.
Even the superposition of a pure dipole and an azimuthal field
is unlikely here, for several reasons.
From a theoretical point of view, this particular combination would 
rule out any ring current.
Moreover, numerical simulations of galactic dynamos always yield
more complex magnetic geometries.
From an observational point of view, if the poloidal field near the GC
were a pure dipole, NRFs crossing the midplane would be curved inward, 
whereas, on the whole, they exhibit a slight outward curvature
\citep{morris_90}.
Besides, the NRF spatial distribution displayed in Fig.~29 of 
\cite{yusef&hc_04} does not at all convey the sense of a pure dipole.
Finally, as noted by \cite{han_02} himself, a global dipole with 
$B_z > 0$ at the position of the Sun should have $B_z < 0$ near the GC,
which is exactly opposite to \cite{novak&crg_03}'s finding.
The matter is in fact a little more subtle, as a dipole field may have 
either sign of $B_z$ near the GC, depending on Galactic polar angle,
$\Theta$ (for reference,
$B_z \begin{array}{c} > \\ \noalign{\vspace{-10pt}} < \end{array} 0$
for $\Theta \begin{array}{c} > \\ \noalign{\vspace{-10pt}} < \end{array} 
\Theta_{\rm crit}$,
where $\Theta_{\rm crit} = \arccos \, (1/\!\sqrt{3}) = 54\fdg7$).
Anyway, a more realistic guess would be that the poloidal field, 
out to beyond the solar circle, is everywhere pointing north and 
slightly curved outward, which would be consistent with the inward 
advection scenario.

The azimuthal component of the GC field, revealed through Faraday
rotation, does not have a well-established parity.
If it has dipolar parity \citep[as suggested by][]{novak&crg_03},
it is probably directly coupled to the dominant poloidal component
and, together with it, connected to the dipolar halo field.
On the other hand, if the azimuthal component has quadrupolar parity
\citep[as suggested by][]{roy&rs_08}, it is probably decoupled from
the poloidal component, but it could connect with the quadrupolar
disk field if the latter is bisymmetric, or at least possesses 
a bisymmetric mode.
If the quadrupolar disk field is axisymmetric or contains 
an axisymmetric mode, the associated poloidal field lines cannot remain 
horizontal all the way in to the GC, but they have to diverge vertically 
somewhere before reaching the GC region.
They can then, on each side of the Galactic plane, turn around in the halo
and arc back to the disk, thereby forming one large or several smaller 
magnetic loops \citep[e.g.,][]{ferriere&s_00}.

To close up the discussion, let us mention the possibility that the GC 
region could harbor its own magnetic system, independent of 
the magnetic field pervading the Galaxy at large.
The origin of such a separate magnetic system could be a GC dynamo, 
possibly modified by outflows from the nucleus.

\subsection{\label{external}Clues from external galaxies}

Could observations of external galaxies shed some light on the properties
of the interstellar magnetic field near the GC?
External galaxies do not give access to 
the wealth of details that can be detected in our own Galaxy, 
but they provide a global view 
which can help fill in some of the gaps in the picture of 
the Galactic magnetic field.

The observational status of interstellar magnetic fields
in external spiral galaxies was recently reviewed by \cite{beck_08}
and \cite{krause_08}.
For practical purposes, external galaxies are generally considered as 
either face-on (if they are not or mildly inclined)
or edge-on (if they are strongly inclined).
While both groups bring along complementary pieces of information,
the tools used to study them are identical and the same as for 
our Galaxy, namely, synchrotron emission and Faraday rotation.

Face-on galaxies are ideally suited to determine the strength 
and the horizontal structure of magnetic fields in galactic disks.
Total field strengths, derived from measurements of the synchrotron total 
intensity together with the equipartition assumption, are $\sim 10~\mu$G
on average -- more precisely, $\sim 5~\mu$G in radio-faint galaxies, 
$\sim 15~\mu$G in more active (higher star-formation rate) galaxies
and up to $\sim (50-100)~\mu$G in starburst galaxies.
Ordered field strengths, derived from measurements of the synchrotron 
polarized intensity, are $\sim (1-5)~\mu$G on average.
The total field is always strongest in the optical spiral arms,
where it can reach up to $\sim (20-30)~\mu$G (for normal galaxies),
whereas the ordered field is generally somewhat stronger in the interarm
regions, where it can reach up to $\sim (10-15)~\mu$G.
Finally, according to synchrotron polarization maps, the ordered field 
tends to follow the orientation of the optical spiral arms,
such that magnetic pitch angles are typically $\sim 10^\circ - 40^\circ$ 
(in absolute value).

Edge-on galaxies are best suited to determine the vertical structure 
of magnetic fields in galactic disks and halos.
Most of them appear to possess extended synchrotron halos, 
the vertical size of which implies that the total field has a vertical 
scale height $\sim 7$~kpc, with little scatter amongst galaxies.
Since the polarization degree increases away from the midplane, 
the ordered field probably has an even greater vertical scale height.
Polarization maps indicate that the ordered field is generally nearly
horizontal close to the midplane.
For galaxies with high-sensitivity measurements (e.g., NGC\,891,
NGC\,5775, NGC\,253, M\,104), the ordered field becomes more vertical 
in the halo, with $|B_z|$ increasing with both $|z|$ and $r$.
The resulting X-shaped field is extremely different from the dipole-like 
field that might be expected from dynamo theory (see Sect.~\ref{connection}), 
not only along the rotation axis, where it goes horizontal instead of 
vertical, but also at large distances from the center, where it diverges 
from the midplane instead of curving back to it.
One possible theoretical explanation for the existence of such X-shaped 
halo fields involves galactic winds 
with roughly radial streamlines
\citep[see, e.g.,][]{dalla&s_08}.

The one notable exception amongst edge-on galaxies is NGC\,4631.
In the disk of this galaxy, the ordered field runs nearly vertical 
throughout the innermost $\sim 5$~kpc,
and it is only at larger radii that it turns roughly horizontal.
In the halo, the ordered field has a more radial appearance, which, 
away from the rotation axis, bears some resemblance to the X shape 
observed in other edge-on galaxies.
The unusual magnetic pattern of NGC\,4631 could perhaps be related to 
the existence of a central starburst and/or to the large-scale rotation 
being almost rigid inside $\sim 5$~kpc.

As in our Galaxy, the measured synchrotron polarization angles only give 
the orientation of the ordered magnetic field, not its direction.
The field direction can sometimes be gathered from 
Faraday rotation measures (of either the galactic synchrotron emission 
itself or background radio sources).
This is how, by combining the observed polarization angles and 
rotation measures, it has been possible to identify well-defined 
azimuthal dynamo modes in a (small) number of galaxies.
For instance, a dominating axisymmetric spiral mode was inferred
in the disks of M\,31 and IC\,342, while a dominating bisymmetric 
spiral mode was suggested for the disk of M\,81. 
Somewhat intriguingly, though, very few field reversals have been 
detected in the disks of external galaxies \citep{beck_01}, 
in contrast to what might be expected from the situation in our Galaxy.
Regarding the vertical parity, evidence was found for a symmetric field 
in the disks of NGC\,891 and IC\,342. 

Observations of external galaxies have a typical spatial resolution 
of a few 100~pc to a few kpc (Marita Krause, private communication).
Evidently, this resolution is way too low to enable detection of
strongly magnetized filaments similar to the NRFs observed near the GC.
Even a large-scale vertical field confined to the innermost $\sim 300$~pc 
would probably escape detection.
In other words, external galaxies could possibly host vertical fields
near their centers, but (with the exception of NGC\,4631)
such vertical fields could not extend beyond the central kpc or so.
Moreover, in none of the edge-on galaxies observed so far
(neither in those featuring an X-shaped field nor in NGC\,4631)
does the halo field resemble a dipole.

Ultimately, what can be retained from observations of external galaxies
is the following:
If our Galaxy does not differ too much from the nearby galaxies whose
magnetic fields have been mapped out, 
one may conclude that the vertical field detected close to the GC 
is only local, i.e., restricted to a region smaller than $\sim 1$~kpc.
One can then imagine two different possibilities:
either the GC field is completely separate from the general Galactic field,
or else it merges smoothly with the poloidal halo field, though not 
in the form of an approximate dipole.

\section{\label{conclu}Conclusions}

Based on all the observational evidence presented in Sect.~\ref{observ},
we may be reasonably confident that the interstellar magnetic field 
in the GC region is approximately poloidal on average in the diffuse
intercloud medium and approximately horizontal in dense interstellar
clouds.
Direct measurements of field strengths are scanty and altogether 
not very informative.
However, our critical discussion of the existing observational and
theoretical estimations in Sect.~\ref{field_strength} prompts us 
to conclude, at least tentatively, 
that the general ISM is pervaded by a relatively weak magnetic field 
($B \sim 10~\mu$G), close to equipartition with cosmic rays
($B_{\rm eq} \sim 10~\mu$G), 
and that it contains a number of localized filamentary structures 
(the observed NRFs) with much stronger fields (up to $B \sim 1$~mG),
clearly above equipartition ($B_{\rm eq} \sim 100~\mu$G).
In dense interstellar clouds, the field would be a few 0.1~mG 
to a few mG strong.

In our view, the high-$B$ filamentary structures are probably
dynamic in nature.
They could, for instance, have a turbulent origin, 
as suggested by \cite{boldyrev&y_06}.
If the GC region is the open magnetic system envisioned by 
these authors, turbulent dynamo action there should produce
strongly magnetized filaments, with $P_{\rm mag} \sim P_{\rm turb}$,
i.e., according to our estimation in Sect.~\ref{field_direction},
$B \sim 0.2$~mG.
Of course, there would be a whole range of field strength,
allowing some filaments to have $B \gtrsim 1$~mG.
Besides, other mechanisms could also come into play, such as
the formation of magnetic wakes behind molecular clouds embedded
in a Galactic wind \citep{shore&l_99}.
The higher ram pressure associated with the wind could then account 
for a larger number of filaments with mG fields.

It is interesting to note that if all NRFs have $B \sim 1$~mG,
the synchrotron data can be reproduced with a roughy uniform 
(within a factor $\sim 10$) relativistic-electron density, $n_{\rm e}$.
Indeed, the factor $\sim 10$ between the equipartition field strengths
inside NRFs ($B_{\rm eq} \sim 100~\mu$G) and in the general ISM 
($B_{\rm eq} \sim 10~\mu$G) translates into a factor $\sim 100$ 
between their respective equipartition electron densities.
Now, since synchrotron emissivity is 
$\propto n_{\rm e} \, B_\perp^{1-\alpha} \, \nu^\alpha$,
the synchrotron emission of NRFs can equally be explained 
by the equipartition values $B \sim B_{\rm eq} \sim 100~\mu$G
and $n_{\rm e} \sim n_{\rm e,eq}$ or by the magnetically dominated values 
$B \sim 1~{\rm mG} \sim 10 \, B_{\rm eq}$ and 
$n_{\rm e} \sim 10^{\alpha-1} \, n_{\rm e,eq}$.
The synchrotron spectral index, $\alpha$, varies considerably amongst NRFs, 
from $\sim -2$ to $\sim +0.4$, with $\alpha \sim -0.6$ being quite typical
\citep[e.g.,][]{reich&wss_88, anantharamaiah&peg_91, gray&nec_95, 
lang&akl_99, lang&me_99, larosa&klh_00, larosa&nlk_04}. 
For NRFs with decreasing spectra ($\alpha < 0$), 
a mG field then implies an electron density $\sim 10 - 1000$ times 
lower than the equipartition value, i.e., within a factor $\sim 10$
of the equipartition electron density in the general ISM.

The situation with highly variable magnetic field and 
roughly uniform relativistic-electron density is in sheer contrast 
with the conventional picture of a uniformly strong magnetic field.
Qualitatively, the former could be understood as the net result of two
antagonistic effects:
On the one hand, when a high-$B$ filament forms by compression,
the attached electrons are compressed together with the field lines.
On the other hand, once the electrons find themselves in a high-$B$
filament, they cool off (via synchrotron radiation) more rapidly 
and they stream away along field lines (at about the Alfv\'en speed) 
faster than electrons in the surrounding medium.

In reality, both the picture of a uniformly strong magnetic field
and the picture of a roughly uniform relativistic-electron density
are probably too extreme.
It is much more likely that the GC region resides in an intermediate state,
where both the field strength and the electron density are higher
in NRFs than in the general ISM.
Such a state is automatically achieved if all NRFs have a field strength
comprised between the equipartition value ($B_{\rm eq} \sim 100~\mu$G)
and $B \sim 1$~mG.

\begin{acknowledgement}{}
The author would like to thank 
J.~Ballet, R.~Beck, S.~Boldyrev, R.~Crutcher, 
M.~Hanasz, J.~ Kn\"odlseder, M.~Krause, A.~Marcowith, F.~Martins, 
I.~Moskalenko, G.~Novak, W.~Reich, A.~Shukurov, A.~Strong, X.~Sun 
and the referee, T.~LaRosa,
for their valuable comments and detailed answers to her questions.
\end{acknowledgement}

\bibliographystyle{aa}
\bibliography{BibTex}

\end{document}